\documentclass[pre]{revtex4-1}
\usepackage{graphicx}
\usepackage{natbib}
\usepackage{bm}
\usepackage{amssymb}
\usepackage{amsmath}
\usepackage{euscript}
\usepackage{esint}
\usepackage{slashed}

\begin{document}

\title{The mean velocity profile of near-wall turbulent flow}

\author{Kirill A. Kazakov}

\affiliation{Department of Theoretical Physics, Physics Faculty, Moscow State
University, 119991, Moscow, Russian Federation}

\begin{abstract}
The issue of analytical derivation of the mean velocity profile in a near-wall turbulent flow is revisited in the context of a two-dimensional channel flow. An approach based on the use of dispersion relations for the flow velocity is developed. It is shown that for an incompressible flow conserving vorticity, there exists a decomposition of the velocity field into rotational and potential components, such that the restriction of the former to an arbitrary cross-section of the channel is a functional of the vorticity and velocity distributions over that cross-section, while the latter is divergence-free and bounded downstream thereof. By eliminating the unknown potential component with the help of a dispersion relation, a nonlinear integro-differential equation for the flow velocity is obtained. It is then analyzed within an asymptotic expansion in the small ratio $v_*/U$ of the friction velocity to the mean flow velocity. Upon statistical averaging in the lowest nontrivial order, this equation relates the mean velocity to the cross-correlation function of the velocity fluctuations. Analysis of the equation reveals existence of two continuous families of solutions, one having the near-wall asymptotic of the form $U \sim \ln^p (y/y_0),$ where $y$ is the distance to the wall, $p>0$ is arbitrary, and the other, $U \sim y^n,$ with $n>0$ also arbitrary except in the limit $n\to 0$ where it turns out to be a universal function of the Reynolds number, $n\sim 1/\ln{\rm Re}.$ It is proved, furthermore, that given a mean velocity distribution having either asymptotic, one can always construct a cross-correlation function so as to satisfy the obtained equation. Next, to bound the strength of the cross-correlations, a condition is introduced that the leading term of the near-wall mean velocity asymptotic be independent of the correlations between velocity fluctuations near the wall and in the bulk. It is found that solutions with the power-law asymptotic do not meet this requirement, whereas those with the logarithmic asymptotic do, $p = 1 + O(\ln{\rm Re}/{\rm Re})$ being uniquely determined by the cross-correlation function. These results are discussed in the light of the existing controversy regarding experimental verification of the law of the wall.
\end{abstract}
\maketitle

\section{Introduction}

Investigation of the near-wall flows constitutes a major part of the turbulence studies. Its results help clarifying basic features of turbulence on the one hand, and find numerous applications in science and industry, on the other. In view of simplicity of their practical realization, the near-wall flows have been extensively studied experimentally, whereas the relative simplicity of the boundary conditions in channels or pipes, and evergrowing computational capabilities allow direct numerical simulation of these flows at still higher Reynolds numbers.\cite{jimenez2013,schultz2013} Wall turbulence is also very attractive from a purely theoretical standpoint, as a system characterized by a rather simple mean flow and well-defined direction of the momentum flow. Moreover, spatial separation of the sources and sinks of momentum, and the well-organized hierarchy of turbulent structures ranged according to their distance to the wall make the study of certain aspects of turbulence even simpler than in the isotropic case.

Despite these simplifications, however, analytical description of the wall turbulence remains nearly as scarce as of other, apparently more complicated flows. The reason for this, shared by all turbulent motions, is the necessity to deal with the nonlinearity of an irregular flow. While construction of an exact general solution of the hydrodynamic equations is now as illusory as it was a hundred years ago, perturbative treatment of the flow nonlinearity has not been successful either. In fact, all attempts to assign ``smallness'' in one way or another to the inertia effects run into fundamental difficulties such as violation of the Galilean invariance (an overview of this issue can be found in Ref.~\cite{frisch1995}). On the other hand, the same nonlinearity is also the reason for incompleteness of the statistical approach, as it requires introduction of a more or less arbitrary model assumption regarding the structure of the correlation functions of the fluctuating flow velocity (the closure problem). In particular, no closed relation following from the basic equations is known that would allow determination of the mean flow velocity profile.

Yet, the above-mentioned specifics of the near-wall turbulent flow has always been a source for the strong belief that its mean velocity profile can be determined indirectly using dimensional analysis and similarity arguments. As is well-known, however, the results of these analyses are not unique: there are two main competing variants of the law of the wall -- the power law, originally due to Prandtl, and von K$\acute{\rm a}$rm$\acute{\rm a}$n's universal logarithmic velocity profile.\cite{karman} Although there are indications that the former is observed in flows characterized by moderately large Reynolds numbers, it is the logarithmic profile that finds overwhelming experimental support as the asymptotic law of the near-wall velocity distribution in the large Reynolds-number limit. Specifically, the recent data\cite{jimenez2013,schultz2013,monty2005,zanoun2009} obtained by direct numerical simulations and highly accurate laboratory measurements in channel flows show that while for ${\rm Re}$ (based on the channel height and bulk mean velocity) up to approximately $60\,000-80\,000$ dependence of the skin-friction coefficient on the Reynolds number is well approximated by a power law, at higher ${\rm Re}$ it is best described by a log law (power-law behavior of the skin friction implies a power law in the mean velocity profile). Things are similar in the inner layers of pipe flows\cite{hultmark2012} and free boundary layers,\cite{degraaff2000} and a pronounced similarity exists between the three types of turbulent flows at very high Reynolds numbers, which goes beyond mere conformity of the mean velocity profiles.\cite{delalamo2004,jimenez2008,monty2009}

It is to be noted, however, that because of the limited measurement accuracy and some vagueness in the spatial extent of the inner layer, the two velocity profiles are not always easy to discern experimentally, so that one and the same data are sometimes interpreted differently. This circumstance on the one hand, and the formal consistency of the incomplete similarity hypothesis underlying the power law, on the other, are the reasons for periodic revivals of theoretical interest to this law, notable recent ones due to Barenblatt\cite{barenblatt1993a} and Oberlack\cite{oberlack2001}. While the former author merely explored consequences of the postulated scaling law for the mean shear, the latter made an attempt to derive the scaling from the Reynolds-averaged Navier-Stokes equation by exploiting its symmetries within the Lie group approach. The conclusion was that both the logarithmic and power law are solutions to this equation, in the sense that they satisfy the Lie group relations. Though this result does not prove actual realizability of either solution, it confirms their consistency from the standpoint of symmetries of the basic equations.

Thus, in spite of the general belief in the logarithmic law, there are empiric indications on the relevance of the power law at Reynolds numbers so large that the turbulence is fully developed, while no theoretical grounds to abandon it in favor of the log-law. This state of affairs is controversial, because it appears to be commonly accepted that there is place for only one law in a fully developed flow. On this account, the fact that the power law ultimately gives way to the log-law as ${\rm Re}$ grows is usually interpreted as a good reason to reject it completely. One of the purposes of the present paper is to make a step towards clarifying this issue.

As was already mentioned, the stumbling block to the development of analytical description of turbulence has been the virtual impossibility to construct, in any form, explicit general solution of the basic hydrodynamic equations. Of course, this difficulty does not pertain to turbulence only, but plagues many other hydrodynamic problems, particularly those involving free boundaries, {\it e.g.,} ablation fronts, condensation discontinuities {\it etc.} A general method to cope with this difficulty was found in the context of premixed flame propagation.\cite{kazakov1,kazakov2} In brief, the method applies to incompressible flows conserving vorticity, and consists in decomposing the flow velocity field into rotational and potential parts in such a way that it becomes possible to explicitly compute the boundary value of the former on some control surface (the flame front, in the case of premixed combustion). The unknown potential component can then be eliminated using a dispersion relation -- a singular-kernel integral equation for its boundary values. The result is a nonlinear integro-differential equation for the flow velocity distribution on the surface. This program will be realized below in the case of turbulent channel flow.

At this point, it is worthwhile to return to the logarithmic law to discuss the underlying hypothesis of complete similarity in some detail. This hypothesis was initially expressed\cite{karman} as a rather strong supposition that the structure of turbulent disturbances in the wall-normal direction can be described infinitesimally in terms of the mean shear and its first derivative in that direction. The subsequent development greatly relaxed the basic assumptions, but one of them has been remaining invariant, namely that the analysis, be it a matching procedure in the overlap region\cite{monin,lumley} or dimensional analysis, is to be carried in terms of the mean shear. The dimensional argument, for instance, goes as follows:\cite{monin,landau} outside the viscous layer, the friction velocity $v_*$ and the distance to the wall $y$ are the only dimensional quantities which the mean shear $dU(y)/dy$ can be built of, so that one can write
\begin{eqnarray}\label{dim1}
\frac{dU}{dy} = \frac{v_*}{\varkappa \,y}\,,
\end{eqnarray}
\noindent $\varkappa$ being a numerical coefficient; integration then yields the logarithmic law for the mean flow velocity $U(y)$
\begin{eqnarray}\label{log}
U = \frac{v_*}{\varkappa}\ln\frac{y}{y_0}\,,
\end{eqnarray}
\noindent where $y_0$ is a viscosity-dependent constant. But applying the same reasoning to a power of the mean velocity, $U^{1/p},$ with $p$ any real number, Eq.~(\ref{dim1}) is replaced by
\begin{eqnarray}\label{dim2}
\frac{dU^{1/p}}{dy} = \frac{v_*^{1/p}}{\varkappa \,y}\,,
\end{eqnarray}
\noindent whence
\begin{eqnarray}\label{logp}
U = \frac{v_*}{\varkappa^p}\ln^p\frac{y}{y_0}\,,
\end{eqnarray}
\noindent instead of Eq.~(\ref{log}). Clearly, the dimensional argument gives no reason to prefer the value $p=1$ to any other, unless one postulates that it is the mean shear that must be independent of viscosity outside the viscous layer. Incidentally, the power law can be derived in the same way (apply the reasoning to $\ln (U/v_*)$).

As observations indicate, the log-scale graphs of $U(y)$ curve upwards in the outer region of the inner layer, therefore, only the values $p\geqslant 1$ in Eq.~(\ref{logp}) can be of practical relevance. The {\it power-log} velocity distributions with $p>1$ are intermediate between the log-law and the power-law distributions, in the sense that for sufficiently large $y$'s they grow faster than the former, but slower than the latter. As will be demonstrated below, all these distributions can in principle represent the near-wall velocity asymptotic, and the answer to the question which one is actually realized depends essentially on the form of the correlation function of the streamwise and wall-normal velocity components.

A recurrent matter in the subsequent analysis is whether or not one apparently natural assumption is satisfied. Its rather loose statement reads:

(A) {\it For a given value of the shear stress, specifics of the outer (core) flow have negligible influence on the structure of the inner layer.}

The real content of this assumption depends on what is meant by the words ``specifics,'' ``influence'' and ``structure,'' but it is clear that in one form or another assumption of this sort need to be introduced in the analysis, at least to justify the notion of the inner layer. The above formulation will be sufficient for the most part of the present study, until a dividing line between the log- and power-laws will have to be drawn. The point of focusing attention on (A) is that by specifying this assumption, it should be possible to obtain a precise formulation of the critical distinction between the two laws. Indeed, this distinction must be related ultimately to the non-universality inherent in the power law, which manifests itself in the dependence of its parameters on the Reynolds number, hence, on the global properties of the flow. On the contrary, the log-law may or may not be universal, but this is immaterial regarding the existence of its limit as ${\rm Re} \to \infty.$ In this connection, the well-known fact is worth recalling that even at very high Reynolds numbers the value of the K$\acute{\rm a}$rm$\acute{\rm a}$n constant $\varkappa$ is actually somewhat different in channel and pipe flows, and that the situation is far less definite regarding the structure of the Reynolds stresses.\cite{monty2009} Identification of the relevant specifics of the outer flow as well as of the conditions under which their influence is negligible is part of the analysis to follow.

This analysis will be carried out for the two-dimensional channel flow. Admittedly, restriction to two dimensions suppresses important features of the turbulent flow, most notably the vortex stretching that drives the energy cascade. Yet there are good reasons to believe in relevance of the present consideration. The point is that it relies not on the properties of the energy cascade, nor even on its existence, but only on those of the cross-correlation function of the velocity fluctuations. Furthermore, the main results turn out to depend rather weakly on the form of this function, being expressed in terms of integrals of its spectral density. At the same time, reduction of dimensionality simplifies the mathematics involved, making particularly convenient the form of the main property of inviscid flow used in the analysis -- conservation of the velocity circulation -- which in two dimensions becomes simply the conservation of vorticity.

The paper is organized as follows. Section~\ref{flowconditions} specifies the channel flow to be studied, displays the governing equations and formulates the boundary conditions in a way which, though somewhat unconventional in the turbulence studies, yet is most adequate to the integral treatment to be given. The main integro-differential equation for the flow fields is derived in Sec.~\ref{derivation}. The derivation begins in Sec.~\ref{decomposition} with a standard representation of the velocity field via the area integral of the vorticity distribution in the channel, which is then gradually simplified with the help of a certain equivalence relation to obtain the rotational component of the flow velocity in the form of an integral of the vorticity distribution over the channel cross-section. The unknown potential component is then projected out in Sec.~\ref{dispersion} using the dispersion relation which expresses analyticity and boundedness of the corresponding complex velocity downstream of that cross-section. Relations of this sort are widely used in various areas of physics, {\it e.g.,} Kramers-Kronig dispersion relations in optics, K\"all\'en-Lehmann spectral formulas in quantum field theory, {\it etc.} In spite of the absence of real dispersion in the present context, for want of a better name, the same term ``dispersion relation'' will be kept for the main integro-differential equation thus obtained, as well as for its descendant derived in Sec.~\ref{asymptoticexp} by expanding in powers of $v_*/U$ and statistical averaging, and further simplified in Sec.~\ref{widechannels} in high channels. A detailed analysis of the dispersion relation is carried out in Secs.~\ref{nearwallasymp}, \ref{reconstruction}, where the velocity profiles discussed above are derived as possible near-wall asymptotics of solutions to the dispersion relation, and the form of the respective cross-correlation functions is established. Throughout the paper, the channel height is assumed large enough to justify the approximations made, and Sec.~\ref{consequences} examines the possibility to go over to the limit of infinite height. For this purpose, the assumption (A) is appropriately specified and applied to the found solutions. The results are discussed in Sec.~\ref{conclusions}. The paper has an appendix that contains derivation of an identity involving the Hilbert operator, used in Sec.~\ref{nearwallasymp} to determine the near-wall mean velocity asymptotic.

\section{Flow conditions}\label{flowconditions}

\begin{figure}
\centering
\includegraphics[width=0.6\textwidth]{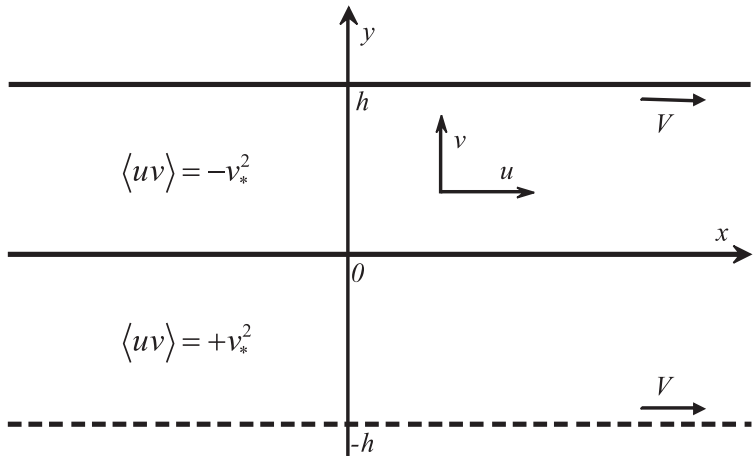}\\
\caption{Schematics of the physical and auxiliary channels.}\label{fig1}
\end{figure}

Consider developed turbulent channel flow between two parallel plates at a distance $h.$ It will be assumed throughout that $h$ is much larger than the viscous length $\nu/v_*$ ($\nu$ is the kinematic fluid viscosity), and that the flow is considered far enough downstream of the channel inlet, so that on average it is steady and homogeneous in the streamwise direction. In regions sufficiently close to either wall (but outside of the viscous layer) the Reynolds shear stress is approximately constant, but if both walls are at rest and the fluid motion is sustained by a pressure gradient, the shear stress is an odd function of the wall-normal coordinate with respect to the channel midpoint. As constancy of the shear stress is the defining property of the near-wall turbulence, its consideration is technically more convenient in the setting where this property holds throughout the channel cross-section. We therefore switch to the picture where one of the channel walls moves in its plane with a constant speed $V,$ the mean pressure gradient in the streamwise direction vanishing, Fig.~\ref{fig1}. The Reynolds number will always be assumed large enough for the turbulence be fully developed, ${\rm Re} = Vh/\nu \gg 1.$ Then the shear stress is constant everywhere in the channel except the regions of width $\sim \nu/v_*$ adjacent to the walls. Denoting $x,y$ the streamwise and wall-normal coordinates, $u,v$ the corresponding flow velocity components, and taking also the fluid of unit density, we thus have $$\langle uv\rangle = - v_*^2,$$ where the angular brackets denote time averaging. Despite formal equality of the walls, when speaking about the near-wall flow we will always refer to the wall at rest, taken to be at $y=0,$ the other wall playing an auxiliary role which is to provide a source of momentum, and to ensure the flow homogeneity. Of course, the above change in the global flow configuration relies on an invariance property of the inner layer, implied by the assumption (A) formulated in the Introduction, but as the subsequent consideration shows, this change is inconsequential in that the achieved simplification of the shear-stress distribution does not preclude existence of any of the velocity profiles of interest -- the log, power, or power-log law.

The velocity field satisfies
\begin{eqnarray}\label{flow1}
\frac{\partial\omega}{\partial t} + v_i\frac{\partial\omega}{\partial x_i} &=& \nu\triangle\omega, \\
\frac{\partial v_i}{\partial x_i} &=& 0,\label{flow2}
\end{eqnarray}
\noindent where $\omega$ is the vorticity,
\begin{eqnarray}\label{vorticity}
\omega = \frac{\partial v}{\partial x} - \frac{\partial u}{\partial y}\,,
\end{eqnarray}
\noindent $(x_1,x_2)\equiv (x,y),$ $(v_1,v_2)\equiv (u,v),$ the subscripts run over $1,2,$ and as usual, summation over repeated indices is understood.

An important premise underlying the present approach is that for the purpose of calculating the mean velocity field, the viscous term in Eq.~(\ref{flow1}) can be neglected outside the viscous layer. As it is part of the somewhat more general standard assumption that this is allowed in the dynamical equations themselves, whose justification can be found in textbooks,\cite{lumley,monin,landau} it will not be discussed further. We only note that it in no way extends over the fields of Reynolds stresses, nor it precludes the possibility of an incomplete similarity. Dependence on viscosity resides in the correlation functions of the velocity fluctuations, and the omission made means merely that their structure cannot be inferred from the inviscid equation. With the right hand side of Eq.~(\ref{flow1}) omitted, it becomes the statement that $\omega$ is constant along the fluid particle trajectories.

Next, to ensure fulfillment of the boundary condition $v=0$ at the wall $y=0,$ it is convenient to introduce an auxiliary channel which is a reflection of the initial one (to be referred to as {\it physical} in what follows) with respect to the wall $y=0.$ Specifically, the new channel has the same height $h,$ occupying the region $y \in (-h,0).$ It is filled with the same fluid as the physical channel, and its wall $y=-h$ moves with the same speed and in the same direction as the wall $y=h,$ Fig.~\ref{fig1}. Unlike the usual formulation of the reflection procedure in the case of laminar flows, however, the instantaneous velocity fields do {\it not} satisfy the relations
\begin{eqnarray}\label{reflection}
u(x,-y) = u(x,y), \quad v(x,-y) = - v(x,-y),
\end{eqnarray}
\noindent which hold only on average. Instead, to formulate the condition of impermeability of the wall $y=0,$ we consider the velocity fields in the regions $y\in(0,h)$ and $y\in(-h,0)$ as restrictions of a unique field describing the flow of the same fluid in the channel $y\in (-h,h)$ without the wall $y=0,$ and require these velocity fields be statistically independent of each other. This formulation of the boundary condition is sufficient for our purposes, and is admissible indeed, for any nonzero flux through the surface $y=0$ (that is, $v(x,0)\ne 0$ for some $x$) would necessarily result in a correlation between the flow velocities to the left and to the right of the surface. As to the other boundary condition at $y=0,$ namely, $u(x,0)=0,$ it does not apply since our consideration has been restricted to the outside of the viscous layer, but the presence of the wall implies that the function $\langle u(x,y)\rangle \equiv U(y),$ obtained as a solution of the inviscid equations, necessarily has a singularity at $y=0$ when formally continued over the viscous layer. Here ``singularity'' means that the function $U(y)$ and/or its derivatives become unbounded in a vicinity of $y=0.$ The presence of this singularity can be taken as a necessary condition on the inviscid solutions, to rule out  solutions proper to the channel $y\in (-h,h),$ such as the trivial solution $v=0,u=V.$ Besides $y=0,$ there can be other singular points of $U(y),$ such as $y=y_0$ in the function (\ref{logp}). All such points must belong to the viscous layer for $U(y)$ to represent a physical velocity distribution. Yet the presence of regions where the function $U(y)$ is formally negative or even complex-valued may be troublesome for the integral relations involving $U(y)$ to be derived below. By this reason, it will be assumed in what follows that whenever $U(y)$ is treated integrally, it is continued over the viscous layer in such a way that it becomes constant within the laminar sublayer, say, $U(y)=v_*,$ the true inviscid solution being smoothly matched to this constant within the buffer layer. Though largely arbitrary, this prescription is justified by the narrowness of the viscous layer, and hence relative smallness of its contribution to the integral quantities.

By the above construction, all statistical properties of the flows in the regions $y\in(0,h)$ and $y\in(-h,0)$ are identical. More precisely, correlation functions $\langle \tilde{u}(x_1,y_1)\cdots v(x_n,y_n)\cdots\rangle$ of the velocity fluctuations are invariant up to a sign under the reflection $(x,y)\to (x,-y),$ the sign being plus or minus depending on whether the number of $v$-components in the product is even or odd. In the case when field derivatives are involved in the product, parity of the correlation function is determined by the total number of $v$-components and $y$-derivatives. In addition to that, statistical properties of the flow in the physical channel are also invariant under the inversion with respect to any point $(x,h/2)$ (or equivalently, under rotation by $\pi$ around this point). In fact, that the flow statistics inherits the symmetries of the boundary conditions is part of the assumption that the turbulent flow is fully developed. This requirement leads to the ``normalization'' condition for the mean flow velocity
\begin{eqnarray}\label{norm}
U\left(h/2\right) = \frac{V}{2}\,,
\end{eqnarray}
\noindent replacing the boundary condition $U(h)=V$ which is not applicable to inviscid solutions.

Of special importance for the following is the cross-correlation function
\begin{eqnarray}\label{crosscor}
R(x_1,y_1;x_2,y_2) = \langle u(x_1,y_1)v(x_2,y_2)\rangle.
\end{eqnarray}
\noindent Since $\langle v \rangle = 0,$ replacing $u$ with the fluctuation $\tilde{u} = u-U$ in this definition makes no difference. Assuming that $R$ is initially defined for positive $y_1,y_2,$ its continuation over the whole channel $y\in (-h,h)$ is, according to the above conditions of statistical independence and parity,
\begin{eqnarray}\label{crosscor1}
R(x_1,y_1;x_2,y_2) = R(x_1,|y_1|;x_2,|y_2|)\left\{\theta(y_1)\theta(y_2) - \theta(-y_1)\theta(-y_2)\right\},
\end{eqnarray}
\noindent where $\theta(x)$ is the step function: $\theta(x)=0$ for $x<0,$ and $\theta(x)=1$ for $x\geqslant 0.$

It remains to guarantee impermeability of the walls at $y=-h$ and $y=h.$ A general way to achieve this is to use an appropriate Green function to represent the velocity field, but since we deal with a straight channel flow in the absence of external fields, it is sufficient to treat it as part of a $2h$-periodic flow obtained by periodic continuation of the flow in the domain $y\in (-h,h)$ along the $y$-axis.

\section{Derivation of the dispersion relation}\label{derivation}

\subsection{Flow decomposition}\label{decomposition}

To extract essential information about the vortical structure of the flow, we gradually simplify its velocity field by stripping it of potential contributions to be collectively denoted $v_i^p,$ with the aim  to express its rotational part $v_i^r = v_i - v_i^p$ in the region downstream of some control surface as a functional of the vorticity distribution over that surface. The simplest choice of the control surface will be made, namely, a vertical cross-section of the flow by the plane $x=0.$ As this procedure has been carried out in detail in more general situations of curved control surfaces and channels of varying height,\cite{kazakov1,kazakov2,jerk1,jerk2,jerk4} it will be reproduced here more compactly using simplifications admitted by the present case.

One starts with the following integral identity which is a consequence of Eq.~(\ref{flow2})
\begin{eqnarray}\label{vint}
v_i &=& \varepsilon_{ik}\partial_k \int\limits_{\Lambda}d
l_l~\varepsilon_{lm} v_m\frac{\ln r}{2\pi} -
\partial_i\int\limits_{\Lambda}d l_k~v_k\frac{\ln r}{2\pi} -
\varepsilon_{ik}\partial_k \int\limits_{\Sigma}d s~\frac{\ln
r}{2\pi}\, \omega \,.
\end{eqnarray}
\noindent Here $\varepsilon_{ik} = - \varepsilon_{ki},\
\varepsilon_{12} = + 1,$ $\partial_i = \partial/\partial x_i;$
$\Sigma$ and $\Lambda$ denote any part of the flow and
its boundary, respectively; $r$ is the distance between an
infinitesimal fluid element $ds$ at the point
$(\tilde{x},\tilde{y})$ and the point of observation
$(x,y)\in \Sigma,$ $r^2 = (x_i - \tilde{x}_i)^2,$ and $d l_i$
is the line element normal to $\Lambda$ and directed outwards from
$\Sigma.$ $\Sigma$ will be taken the rectangular set of points with $y\in [-R,R]$ between the lines $x=0$ and $x=R,$ where $R\gg h$ is meant to go ultimately to infinity, so that $\Sigma$ will occupy the whole region downstream of $x=0.$

\begin{figure}
\centering
\includegraphics[width=0.8\textwidth]{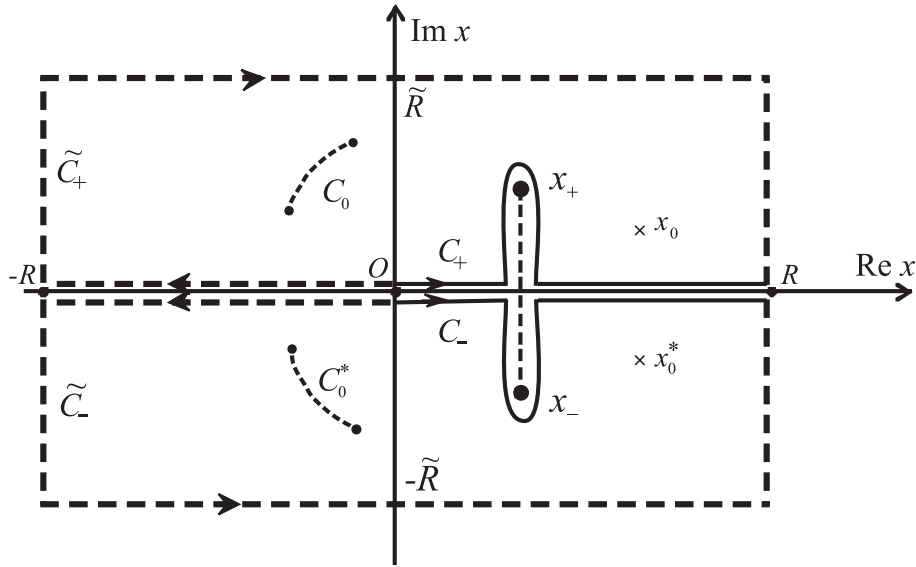}
\caption{Singularities of the integrand in Eq.~(\ref{vint2}), and contours of integration over $\tilde{x}$ used to extract the singular contribution to the rotational velocity component. $x_{\pm}$ are the branch points of the logarithm, $x_0,$ $C_0$ are a pole and a cut connecting branch points of the function $\omega[0,\eta(\tilde{x},\tilde{y},t),\tau(\tilde{x},\tilde{y},t)].$}\label{fig2}
\end{figure}

Two fields $f_i(x,y),$ $\tilde{f}_i(x,y)$ are said equivalent, $f_i(x,y)\stackrel{\circ}{=}\tilde{f}_i(x,y),$ if $\phi_i(x,y) = f_i(x,y)-\tilde{f}_i(x,y)$ satisfies $\partial_iD\phi_i = 0,$ $\varepsilon_{ik}\partial_iD\phi_k=0,$ and $D\phi_i$ is bounded in $\Sigma,$ where $D\phi_i$ denotes any first-order spatial derivative of $\phi_i.$ This extra differentiation $D$ and the condition of boundedness will be found important when taking the limit $R\to \infty$ and applying a dispersion relation to the potential part of the velocity field. In particular, $D$-differentiation makes the line integrals on the right of Eq.~(\ref{vint}) convergent in this limit, and hence the corresponding contributions bounded. Since these terms are also irrotational and divergence-free, one has
\begin{eqnarray}
v_i \stackrel{\circ}{=} -
\varepsilon_{ik}\partial_k \int\limits_{\Sigma}d s~\frac{\ln
r}{2\pi}\, \omega \,,\nonumber
\end{eqnarray}
\noindent or with all the field arguments displayed,
\begin{eqnarray}\label{vint1}
v_i(x,y,t) \stackrel{\circ}{=} -
\varepsilon_{ik}\partial_k \int\limits_{\Sigma}d\tilde{x}d\tilde{y}~\frac{\ln
r}{2\pi}\, \omega(\tilde{x},\tilde{y},t) \,.
\end{eqnarray}
\noindent The purpose of the subsequent equivalence transformation is to reduce the right hand side of Eq.~(\ref{vint1}) to a one-dimensional integral over $\tilde{y}.$ It uses deformation of the integration contour to extract the singularity of the integral kernel at $r=0.$ It is this singularity that determines vortical structure of the flow in a vicinity of the given observation point, because contributions of the fluid elements lying outside of the vicinity are irrotational inside. Since the control surface $(x=0)$ is planar, there is no need to switch to the Lagrangian time of the fluid elements as was done in Refs.~\cite{jerk2,jerk4}, and the contour deformation can be performed directly in terms of the $\tilde{x}$-integration variable. Let $X(\eta,\tau,t),$ $Y(\eta,\tau,t)$ denote the $x$- and $y$-coordinates at the current time $t$ of the fluid element that crossed the point $(0,\eta)$ at the instant $\tau.$ Then the integral appearing in Eq.~(\ref{vint1}) can be written, on account of the vorticity conservation, as
\begin{eqnarray}\label{vint2}
\int\limits_{\Sigma}d s~\omega\ln
r = \int\limits_{-R}^{R}d\tilde{y}\int\limits_{0}^{R}d\tilde{x}
\,\omega[0,\eta(\tilde{x},\tilde{y},t),\tau(\tilde{x},\tilde{y},t)]
\ln\left\{(x-\tilde{x})^2+(y-\tilde{y})^2\right\}^{1/2}\,,
\end{eqnarray}
\noindent where $\{\eta(\tilde{x},\tilde{y},t),$ $\tau(\tilde{x},\tilde{y},t)\}$ is the solution of the equations
\begin{eqnarray}\label{trajectory}
X(\eta,\tau,t) = \tilde{x}, \quad Y(\eta,\tau,t) = \tilde{y}.
\end{eqnarray}
\noindent This solution exists for any point $(\tilde{x},\tilde{y}),$ except the zero-measure set $\tilde{y}=2hn,$ $n\in Z,$ because the mean fluid velocity is positive outside of this set. It is also unique, because different trajectories do not intersect. Singularities of the integrand, considered as a function of the complex variable $\tilde{x},$ are the logarithm branch points
\begin{eqnarray}\label{branch}
x_+ = x + i|y-\tilde{y}|, \quad x_- = x - i|y-\tilde{y}|,
\end{eqnarray}
\noindent and those corresponding to the function $\omega[0,\eta(\tilde{x},\tilde{y},t),\tau(\tilde{x},\tilde{y},t)].$ The latter are necessarily present as $\omega[0,\eta(\tilde{x},\tilde{y},t),\tau(\tilde{x},\tilde{y},t)]$ is not constant, and are expected to be located at distances $\lesssim h$ from the real axis, because the largest eddy size is $h.$ It is not difficult to see that the contributions of such singularities are potential. Consider the field
$$\phi_i = -\frac{1}{2\pi}\omega[0,\eta(\tilde{x},\tilde{y},t),\tau(\tilde{x},\tilde{y},t)] \varepsilon_{ik}\partial_k\ln\left\{(x-\tilde{x})^2+(y-\tilde{y})^2\right\}^{1/2}\,,$$ which on account of Eq.~(\ref{vint2}) represents the integrand in Eq.~(\ref{vint1}). It satisfies $\partial_i\phi_i \equiv 0,$ and $\varepsilon_{ik}\partial_i\phi_k = 0$ for $(x,y) \ne (\tilde{x},\tilde{y}),$ implying that it is potential for any $\tilde{x}$ with a non-zero imaginary part. At the same time, the function $\omega[0,\eta(\tilde{x},\tilde{y},t),\tau(\tilde{x},\tilde{y},t)]$ cannot be singular at real $\tilde{x},$ because $\eta(\tilde{x},\tilde{y},t)$ and $\tau(\tilde{x},\tilde{y},t)$ are real when $\tilde{x}$ is real, whereas all the flow functions involved are regular for real values of their arguments. Therefore, contribution of any singularity of $\omega[0,\eta(\tilde{x},\tilde{y},t),\tau(\tilde{x},\tilde{y},t)],$ encountered when deforming the contour of $\tilde{x}$-integration, such as a pole $x_0$ of arbitrary order or a cut $C_0$ connecting branch points, sketched in Fig.~\ref{fig2}, is potential indeed. Also, since each such singularity can be surrounded by a contour of finite length, the corresponding terms in $Dv_i$ are bounded, and therefore, equivalent to zero. In other words, all such contributions can be omitted altogether. On the other hand, the points $x_{\pm}$ tend to the real axis as $\tilde{y}\to y,$ clipping the integration contour (the segment $\tilde{x}\in [0,R]$). It is this singularity that determines the vortical structure of the flow. To extract its contribution, we replace the $\tilde{x}$-integral over $[0,R]$ by one half of the integral over the contour $C=C_+\cup C_-$ shown in Fig.~\ref{fig2}, compensating the phase change of $r=\sqrt{(x - \tilde{x})^2 + (y - \tilde{y})^2}$ by adding an integral of the jump of the logarithm at the cut connecting the points $x_-,$ $x_+.$ The contour $C$ can then be moved away from $x_{\pm}$ by deforming it into the contour $\tilde{C} = \tilde{C}_+\cup \tilde{C}_-.$ To embrace $x_{\pm}$ for all $y,\tilde{y}\in \Sigma,$ the outer (horizontal) segments of $\tilde{C}$ should be at a distance $\tilde{R}>2R$ from the real axis.

We thus obtain
\begin{eqnarray}\label{vint3}
v_i \stackrel{\circ}{=} \frac{i}{4}\varepsilon_{ik}\partial_k \int\limits_{-R}^{R}d\tilde{y}\int\limits_{x_-}^{x_+}d\tilde{x}\,
\omega[0,\eta(\tilde{x},\tilde{y},t),\tau(\tilde{x},\tilde{y},t)] + \int\limits_{-R}^{R}d\tilde{y}\int\limits_{\tilde{C}}d\tilde{x}\,\phi_i.
\end{eqnarray}
\noindent The last term here is potential in $\Sigma,$ but it is difficult to give a general proof that it remains bounded in the limit $R\to \infty.$ If it does, then so does the first term, and vice versa, because the left hand side is bounded as representing the physical field (more precisely, this holds for the corresponding derivatives, according to the definition of  $\stackrel{\circ}{=}$). In this case the last term $\stackrel{\circ}{=} 0,$ and can be omitted. In other words, the problem is that despite seemingly different structure of the two integrals, a possibility cannot be excluded that both are divergent in the limit $R\to \infty,$ the divergences vanishing only in their sum. A general way to overcome this difficulty is to ``reshuffle'' possible divergences by introducing an intermediate regularization of the integrals that would ensure separate existence of their limits as $R\to \infty,$ and then remove the regularization. The most convenient practical recipe is the following. We observe that since the integrand in the first term is independent of $x,y,$ by virtue of the Newton-Leibniz theorem it is a one-dimensional integral over $\tilde{y}$ of a function defined on the control surface. By the construction, the functions $\eta(\tilde{x},\tilde{y},t),$ $\tau(\tilde{x},\tilde{y},t)$ are $2h$-periodic with respect to $\tilde{y},$ hence, so is the function $\omega[0,\eta(\tilde{x},\tilde{y},t),\tau(\tilde{x},\tilde{y},t)].$ Therefore, the first integral in Eq.~(\ref{vint3}) would be periodic for $R = \infty,$ if finite. An appropriate regularization of such integrals is the introduction of an exponential damping $e^{-\mu|y-\tilde{y}|}$ into integrands, where the parameter $\mu$ is sufficiently large. One can then go over to the limit $R\to \infty,$ so that Eq.~(\ref{vint3}) is replaced by
\begin{eqnarray}\label{vint4}
v_i \stackrel{\circ}{=} \frac{i}{4}\varepsilon_{ik}\partial_k \int\limits_{-\infty}^{\infty}d\tilde{y}e^{-\mu|y-\tilde{y}|}\int\limits_{x_-}^{x_+}d\tilde{x}\,
\omega[0,\eta(\tilde{x},\tilde{y},t),\tau(\tilde{x},\tilde{y},t)] + \int\limits_{-\infty}^{\infty}d\tilde{y}e^{-\mu|y-\tilde{y}|}
\int\limits_{\tilde{C}}d\tilde{x}\,\phi_i.
\end{eqnarray}
\noindent Now, the first integral can be analytically continued in $\mu$ to $\mu = 0,$ as this continuation exists in view of the above-mentioned $2h$-periodicity of the integral. On the other hand, for $\mu = 0$ the right hand side of Eq.~(\ref{vint4}) coincides with that of Eq.~(\ref{vint3}), so that up to a constant, it defines the same velocity field. Moreover, it can be proved\cite{jerk2} that the procedure just described preserves potentiality of the second term in Eq.~(\ref{vint4}). This equation thus reduces to
\begin{eqnarray}\label{vint5}
v_i \stackrel{\circ}{=} \left.\frac{i}{4}\varepsilon_{ik}\partial_k \int\limits_{-\infty}^{\infty}d\tilde{y}e^{-\mu|y-\tilde{y}|}\int\limits_{x_-}^{x_+}d\tilde{x}\,
\omega[0,\eta(\tilde{x},\tilde{y},t),\tau(\tilde{x},\tilde{y},t)] \right|_{\mu=0}.
\end{eqnarray}\noindent Concrete realizations of this procedure in application to various problems of premixed combustion can be found in Refs.~\cite{jerk2,jerk3,jerk4}

Fortunately, the case under consideration admits an important simplification that makes the use of analytic continuation superfluous. Namely, it will be shown in Sec.~\ref{asymptoticexp} that the obtained expression for the rotational velocity component turns out to be localizable, that is, at each order of an asymptotic expansion with respect to $v_*/U,$ a finite-order derivative of the integrand in Eq.~(\ref{vint5}) with respect to $y$ vanishes for $\mu=0$ and $\tilde{y}\ne y.$ A rigorous consideration\cite{jerk4} shows that the result of formal differentiation of a localizable expression can differ from that obtained using the analytic continuation by a term which is position-dependent for general (curved) control surfaces, but in the case of a plane this term is constant, equal to the mean of the expression over that plane. Therefore, by switching to a $y$-derivative of sufficiently high order, such that the obtained expression is explicitly finite and has zero mean, one can ensure correctness of the result. As this rule greatly simplifies calculations, we use it below omitting the factor $e^{-\mu|y-\tilde{y}|}$ and the accompanying symbol of analytic continuation to $\mu=0$ in Eq.~(\ref{vint5}).

\subsection{The dispersion relation}\label{dispersion}

Given a potential velocity field $v^p_i,$ the complex combination $v^p_1-iv^p_2 = u^p-iv^p \equiv g^p$ is an analytical function of the complex variable $x+iy\equiv z.$ If, in addition to that, it is periodic in $y$ and bounded in the domain $x\geqslant 0,$ its restriction to the control surface $(x=0)$ satisfies the relation
\begin{eqnarray}\label{cauchy}
(1-i\hat{H})\left(g^p\right)' = 0,
\end{eqnarray}\noindent where prime denotes differentiation with respect to $y,$ and $\hat{H}$ is the Hilbert operator defined on functions $a(y)$ with zero mean over $y\in (-\infty,\infty)$ by
\begin{eqnarray}\label{hilbertinf}
(\hat{H}a)(y) = \frac{1}{\pi}\fint\limits_{-\infty}^{\infty}d\tilde{y}\frac{a(\tilde{y})}{\tilde{y}-y}\,.
\end{eqnarray}\noindent Equation (\ref{cauchy}) is a consequence of the Cauchy theorem. For a $2h$-periodic function $a(y),$ the definition of $\hat{H}$ can be rewritten, in view of the well-known expansion of the cotangent, as
\begin{eqnarray}\label{hilbert}
(\hat{H}a)(y) = \frac{1}{2h}\fint\limits_{-h}^{h}d\tilde{y}a(\tilde{y})
\cot\left\{\frac{\pi}{2h}(\tilde{y}-y)\right\}.
\end{eqnarray}\noindent The relation (\ref{cauchy}) holds for higher derivatives $d^ng^p/dy^n$ as well, because $d^n g^p/dz^n$ satisfies the requirements imposed on $g^p.$

According to Eq.~(\ref{vint5}), these requirements are also met by the field
\begin{eqnarray}
g(x,y) - \frac{i}{4} \int\limits_{-\infty}^{\infty}d\tilde{y}\,
\left\{\omega[0,\eta(x_+,\tilde{y},t),\tau(x_+,\tilde{y},t)]\slashed{\partial}x_+ - \omega[0,\eta(x_-,\tilde{y},t),\tau(x_-,\tilde{y},t)]\slashed{\partial}x_-\right\}, \nonumber
\end{eqnarray}\noindent where $g = u-iv,$ $\slashed{\partial}\equiv \partial/\partial y + i\partial/\partial x.$ Therefore, substituting $\slashed{\partial}x_{\pm} = i \pm i\chi(y-\tilde{y}),$ where $\chi(y)$ is the sign function,
$$\chi(y) = \left\{
\begin{array}{cc}
+1,& y>0\,,\\
\phantom{+}0, & y = 0\,,\\
-1,&  y<0\,,
\end{array}
\right.$$ we find
\begin{eqnarray}\label{main}
(1-i\hat{H})\left[ u - iv + \frac{1}{4} \int\limits_{-\infty}^{\infty}d\tilde{y}
\left\{\phantom{\int}\hspace{-0,4cm}\omega[\eta(i|y-\tilde{y}|,\tilde{y},t),\tau(i|y-\tilde{y}|,\tilde{y},t)][1 + \chi(y-\tilde{y})] \nonumber\right.\right.\\ \left.\left.  - \omega[\eta(-i|y-\tilde{y}|,\tilde{y},t),\tau(-i|y-\tilde{y}|,\tilde{y},t)][1 - \chi(y-\tilde{y})]\hspace{-0,4cm}\phantom{\int}\right\}\hspace{-0,6cm}\phantom{\int\limits_{-\infty}^{\infty}}\right]' =
0.
\end{eqnarray}\noindent Hereon, $u,v,\omega$ denote restrictions of the fields $u(x,y,t),$ $v(x,y,t),$ $\omega(x,y,t)$ to the control surface, $u(y,t)\equiv u(0,y,t),$ {\it etc.}

Next, we recall that the the function $\eta(\tilde{x},\tilde{y},t)$ is defined by Eq.~(\ref{trajectory}) wherein trajectories $(X(\eta,\tau,t),Y(\eta,\tau,t))$ cross the control surface for $t=\tau,$ whence, $X(\eta,\tau,\tau) = 0,$ $Y(\eta,\tau,\tau) = \eta.$  On the other hand, we know that the vortical structure near given observation point $(x,y)$ is determined by an infinitesimal vicinity of the point $(0,\eta(x,y,t))$ on the control surface. Therefore, the integrand in Eq.~(\ref{main}) can be calculated using the following approximation of trajectories, valid for infinitesimal $(t-\tau)$:
\begin{eqnarray}\label{trajectory1}
X(\eta,\tau,t) = (t-\tau)u(\eta,t), \quad Y(\eta,\tau,t) = \eta + (t-\tau)v(\eta,t).
\end{eqnarray}\noindent Substituting this into Eq.~(\ref{trajectory}) yields an equation for the function $\eta(x_{\pm},\tilde{y},t)$
\begin{eqnarray}\label{eta}
\eta + x_{\pm}\,\frac{v(\eta,t)}{u(\eta,t)} = \tilde{y}.
\end{eqnarray}\noindent Once $\eta(x_{\pm},\tilde{y},t)$ is found, the function $\tau(x_{\pm},\tilde{y},t)$ is given by
\begin{eqnarray}\label{tau}
\tau(x_{\pm},\tilde{y},t) = t - \frac{x_{\pm}}{u[\eta(x_{\pm},\tilde{y},t),t]}\,.
\end{eqnarray}\noindent

Despite Eq.~(\ref{main}) is complex, only its real part represents an independent relation. This is because its left hand side vanishes identically under the action of $(1+i\hat{H})$ by virtue of the well-known property of the Hilbert operator, $\hat{H}^2 = -1.$ Together with Eq.~(\ref{eta}), it thus yields an integro-differential relation for the functions $u(\eta,t),$ $v(\eta,t).$

\subsection{Asymptotic expansion of the dispersion relation}\label{asymptoticexp}

Equation (\ref{main}) is generally rather complicated, but it can be greatly simplified in two respects. One is that the condition $v_*/V\ll 1$ allows asymptotic expansion of Eqs.~(\ref{main}), (\ref{eta}) with respect to $v/U,$ $\tilde{u}/U,$ because $v,\tilde{u}=O(v_*),$ and $v_*/U\ll 1$ holds outside the viscous layer ($\tilde{u}$ is the fluctuation of the streamwise  velocity component, $\tilde{u}= u - U$). This expansion will be carried out below to the second order which is the leading nontrivial order, that is, the one where Eq.~(\ref{main}) first becomes a nontrivial relation upon averaging over time. The other simplification is the disregard of the memory effects associated with the non-locality in time of the integrand in Eq.~(\ref{main}). It is not difficult to show that the presence of the coordinate dependence in the argument $\tau$ of the function $\omega[\eta(x_+,\tilde{y},t),\tau(x_+,\tilde{y},t)]$ does not affect the vorticity distribution. Moreover, the divergenceless of the vorticity component would neither be violated if $\tau(x_+,\tilde{y},t)$ is replaced by $t.$ However, the boundedness property is not generally preserved, so that the replacement $\tau(x_+,\tilde{y},t)\to t$ in $\omega$ is not an equivalence transformation in the sense of $\stackrel{\circ}{=}.$ In this connection, it is perhaps worth to invoke the experience gained from the theory of premixed flame propagation. Discarding the memory effects does not change the general structure of the Darrieus-Landau dispersion relation which describes unstable behavior of nearly planar flame fronts, though it does change numerical coefficients therein, and this change turns out to be important in the limit of small gas expansion where the front dynamics becomes slow.\cite{jerk2} In general, the relative role of the memory effects decreases together with the characteristic time of the process, and so neglecting them makes sense in a sufficiently fast flow. Since $[\tau(x_+,\tilde{y},t) - t]\sim 1/V$ according to Eq.~(\ref{tau}), this formally means that the memory effects are treated in the zeroth order approximation with respect to $v_*/V.$ Having in mind to use Eq.~(\ref{main}) averaged over time, it is also possible that validity of the replacement $\tau(x_+,\tilde{y},t)\to t$ can be justified beyond this approximation, but for now it will be adopted as a plausible assumption.

We begin with solving Eq.~(\ref{eta}) approximately. Omitting the time argument for brevity, the first-order solution $\eta(x_{\pm},\tilde{y},t) \equiv \eta_{\pm}$ of this equation reads
\begin{eqnarray}\label{eta1}
\eta_{\pm} = \tilde{y} - x_{\pm}\,\frac{v(\tilde{y})}{U(\tilde{y})}.
\end{eqnarray}\noindent Substituting this in the small second term on the left of Eq.~(\ref{eta}), and expanding again gives
\begin{eqnarray}\label{eta2}
\eta_{\pm} = \tilde{y} - x_{\pm}\left[\frac{v(\tilde{y})}{U(\tilde{y})} - \frac{v(\tilde{y})\tilde{u}(\tilde{y}) + x_{\pm} v(\tilde{y})v'(\tilde{y})}{U^2(\tilde{y})}\right].
\end{eqnarray}\noindent In all these formulas, $x_{\pm}=\pm i|y-\tilde{y}|.$ This expression is then used to expand the functions $\omega(\eta_{\pm})$ in the integrand of Eq.~(\ref{main})
$$\omega(\eta_{\pm}) = \omega(\tilde{y}) - \omega'(\tilde{y})x_{\pm}\left[\frac{v(\tilde{y})}{U(\tilde{y})} - \frac{v(\tilde{y})\tilde{u}(\tilde{y}) + x_{\pm} v(\tilde{y})v'(\tilde{y})}{U^2(\tilde{y})}\right] + \frac{x^2_{\pm}}{2}\omega''(\tilde{y})\frac{v^2(\tilde{y})}{U^2(\tilde{y})}\,.$$
Inserting this into Eq.~(\ref{main}), extracting its real part, and recalling the definition (\ref{vorticity}) of $\omega$ yields, after some simple algebra,
\begin{eqnarray}
\hat{H}\frac{\partial v}{\partial y} - \frac{\partial v}{\partial x} + \frac{\hat{H}}{2}\int\limits_{-\infty}^{\infty}d\tilde{y}\chi(y-\tilde{y})\omega'(\tilde{y})\left\{ \frac{v(\tilde{y})}{U(\tilde{y})} - \frac{v(\tilde{y})\tilde{u}(\tilde{y})}{U^2(\tilde{y})}\right\} + \int\limits_{-\infty}^{\infty}d\tilde{y}|y-\tilde{y}|
\frac{\left[\omega'(\tilde{y})v^2(\tilde{y})\right]'}{2U^2(\tilde{y})} = 0, \nonumber
\end{eqnarray}\noindent where $\partial v/\partial x$ stands for $[\partial v(x,y)/\partial x]_{x=0}.$ As was explained at the end of Sec.~\ref{decomposition}, the unregularized  $\tilde{y}$-integrals here have only formal meaning. But the second $y$-derivative of this equation is already of strict meaning,
\begin{eqnarray}\label{main1}
\frac{\partial}{\partial y}\left[\hat{H}\frac{\partial^2 v}{\partial y^2} - \frac{\partial^2 v}{\partial y\partial x}\right] + \hat{H}\frac{\partial}{\partial y}\left(\frac{\omega'v}{U}\right) - \hat{H}\frac{\partial}{\partial y}\left(\frac{\omega'v\tilde{u}}{U^2}\right) +
\left[\frac{\left(\omega'v^2\right)'}{U^2} - C_0 \right]= 0,
\end{eqnarray}\noindent where $$C_0 = \frac{1}{2h}\int\limits_{-h}^{h}dy\frac{\left(\omega'v^2\right)'}{U^2}\,.$$ Indeed, all terms in this equation are explicitly finite and have  zero mean over the control surface [or, which is the same, over $y\in (-h,h)].$ We note for the future that $\langle C_0\rangle \equiv 0,$ because $\langle \left(\omega'v^2\right)'\rangle$ is an odd function of $y$ (Cf. account of parity properties at the end of Sec.~\ref{flowconditions}).

It is important that Eq.~(\ref{main1}) can be rewritten, within the second-order approximation, in a form that in Sec.~\ref{widechannels} will be used to resolve the closure problem in the context of the present approach. To this end, consider the first-order part of Eq.~(\ref{main1}) which integrates to
\begin{eqnarray}\label{firstorder}
\hat{H}\frac{\partial^2 v}{\partial y^2} - \frac{\partial^2 v}{\partial y\partial x} + \hat{H}\left(\frac{\omega'v}{U} - C\right) = 0,
\end{eqnarray}\noindent the integration constant $C=C(t)$ being the spatial average of $\omega'v/U$ over $y\in (-h,h).$ Acting on this equation by $\hat{H},$ using the identity $\hat{H}^2 = -1,$ and multiplying the result by $Uv$ or $U\tilde{u}$ gives, respectively,
\begin{eqnarray}
\omega' v^2 &=& CUv - Uv\hat{H}\frac{\partial^2 v}{\partial y\partial x} - U v \frac{\partial^2 v}{\partial y^2}\,,\\
\omega' v\tilde{u} &=& CU\tilde{u} - U\tilde{u}\hat{H}\frac{\partial^2 v}{\partial y\partial x} - U \tilde{u} \frac{\partial^2 v}{\partial y^2}\,.
\end{eqnarray}\noindent Putting this into Eq.~(\ref{main1}), and averaging over time yields
\begin{eqnarray}
\hat{H}\frac{d}{dy}\frac{1}{U}\left\langle \tilde{u}\hat{H}\frac{\partial^2 v}{\partial y\partial x} + v\frac{\partial^2 v}{\partial y\partial x} + \tilde{u}\frac{\partial^2 v}{\partial y^2} - v\frac{\partial^2 \tilde{u}}{\partial y^2} - C\tilde{u}\right\rangle = \frac{1}{U^2}\frac{d}{d y}\,U\!\left\langle v\hat{H}\frac{\partial^2v}{\partial y\partial x} + v\frac{\partial^2 v}{\partial y^2} - C v\right\rangle. \nonumber
\end{eqnarray}\noindent As the time-averaged quantities defined on the control surface depend only on $y,$ the partial $y$-derivatives acting thereupon have been replaced by $d/dy.$ Finally, we note that
\begin{eqnarray}\label{avident}
\left\langle\tilde{u}\frac{\partial^2 v}{\partial y^2}\right\rangle = \left\langle v\frac{\partial^2 \tilde{u}}{\partial y^2}\right\rangle.
\end{eqnarray}
\noindent Indeed, taking the time-average of Eq.~(\ref{flow1}),
\begin{eqnarray}
\left\langle v_i\frac{\partial\omega}{\partial x_i}\right\rangle = 0,\nonumber
\end{eqnarray}
\noindent inserting the definition of $\omega,$ and using Eq.~(\ref{flow2}) gives
\begin{eqnarray}
\left\langle \tilde{u}\frac{\partial^2 v}{\partial x^2} - v\frac{\partial^2 \tilde{u}}{\partial x^2} + \tilde{u}\frac{\partial^2 v}{\partial y^2} - v\frac{\partial^2 \tilde{u}}{\partial y^2}\right\rangle = 0.\nonumber
\end{eqnarray}
\noindent But in view of the averaged flow homogeneity in the streamwise direction, the cross-correlation function (\ref{crosscor}) depends on $x_1,x_2$ only through the difference $(x_1-x_2),$ therefore,
$$ \left\langle \tilde{u}\frac{\partial^2 v}{\partial x^2}\right\rangle = \left\langle v\frac{\partial^2 \tilde{u}}{\partial x^2} \right\rangle,$$ which proves the identity (\ref{avident}). We thus arrive at the following equation
\begin{eqnarray}\label{main2}
\hat{H}\frac{d}{dy}\frac{1}{U}\left\langle \tilde{u}\hat{H}\frac{\partial^2 v}{\partial y\partial x} + v\frac{\partial^2 v}{\partial y\partial x} - C\tilde{u}\right\rangle = \frac{1}{U^2}\frac{d}{d y}\,U\!\left\langle v\hat{H}\frac{\partial^2v}{\partial y\partial x} + v\frac{\partial^2 v}{\partial y^2} - C v\right\rangle.
\end{eqnarray}\noindent

Detailed analysis of this equation will be carried out in the next section, but before it  several important points of its derivation are to be discussed. First,
the transformation just performed turned out to be possible because of the special feature of Eq.~(\ref{main1}) that the terms proportional to $\omega''v^2$ and $\omega'(v^2)'$ had combined into the last term in Eq.~(\ref{main1}), which involves only the product $\omega'v^2.$ As a result, triple correlations of fluctuating fields now enter the equation only spatially averaged, through the terms $\langle C\tilde{u}\rangle,$ $\langle Cv\rangle,$ where $C = (1/2h)\int_{-h}^{h}dy\omega'v/U.$ Second, it is to be noted that in the course of asymptotic expansion of Eq.~(\ref{main}) the flow velocity itself  has not been expanded (neither $U,$ nor $\tilde{u},v$). In other words, solutions to this equations are not treated as asymptotic series. In fact, doing so would be inconsistent, as can be seen from Eq.~(\ref{firstorder}): an attempt to write $v_i = v^{(0)}_i + v^{(1)}_i + \cdots,$ where $v^{(1)}/v^{(0)} = O(v_*/U),$ would give
\begin{eqnarray}
\hat{H}\frac{\partial^2 v^{(0)}}{\partial y^2} - \frac{\partial^2 v^{(0)}}{\partial y\partial x} &=& 0, \nonumber\\
\hat{H}\frac{\partial^2 v^{(1)}}{\partial y^2} - \frac{\partial^2 v^{(1)}}{\partial y\partial x} + \hat{H}\left(\frac{\left[\omega^{(0)}\right]'v^{(0)}}{U^{(0)}} - C\right) &=& 0.\nonumber
\end{eqnarray}\noindent But the first of these equations means that the field $v^{(0)}_i$ is potential, whence $\omega^{(0)} = 0,$ and so the second equation fails to couple $v^{(1)}$ to $v^{(0)}$ and $U,$ merely duplicating the first. It is also worth noting that even if the expansion of $v_i$ was consistent, it would actually be inadequate to the main purpose of Eq.~(\ref{main2}), which is to provide a relation between measurable quantities (mean velocity and velocity correlations). Therefore, using the expansion $v_i = v^{(0)}_i + v^{(1)}_i + \cdots$ would result in a sequence of relations involving quantities such as $\langle v^{(0)''}\tilde{u}^{(1)}\rangle,$ which could be determined only by solving the complete system of hydrodynamic equations, a task the present approach is designed to overcome.

Incidentally, it is not difficult to see that in the first-order approximation, there is no turbulence driven by shear. In fact, averaging Eq.~(\ref{firstorder}) over time and using $\langle v\rangle \equiv 0,$ $\langle C\rangle \equiv 0$ (the latter follows from the oddness of the combination $\langle\omega'v\rangle/U = - \langle v\triangle u\rangle/U$), one finds $\hat{H}\left(\langle v\triangle \tilde{u}\rangle/U\right) = 0,$ hence, $\langle v\triangle \tilde{u}\rangle = 0,$ which for a realistic cross-correlation function means $\langle v \tilde{u}\rangle = 0,$ that is $v_*=0.$ A nontrivial shear turbulence appears in the next, second-order approximation, wherein the cross-correlation function is not zero, but is related to $U$ by Eq.~(\ref{main2}).

At last, it is quite clear that the above procedure of asymptotic expansion of Eq.~(\ref{main}) can be carried out to any order desired, because the integrals $\int_{-\infty}^{\infty}d\tilde{y}
\omega[\eta(x_{\pm},\tilde{y},t)][1 \pm \chi(y-\tilde{y})]$ are localizable at any finite order.

\section{Analysis of the dispersion relation}\label{analysis}

\subsection{Equation~(\ref{main2}) in high channels}\label{widechannels}

Equation (\ref{main2}) relates the mean flow velocity $U(y),$ the cross-correlation function (\ref{crosscor}), and triple correlations of velocity fluctuations. The presence of the latter constitutes a closure problem: in general, additional information regarding the structure of triple correlations is needed to make of Eq.~(\ref{main2}) a useful relation. It is important, however, that the triple products of field fluctuations enter this equation only in the combinations $\langle C\tilde{u}\rangle,$ $\langle Cv\rangle,$ where
$$C = \frac{1}{2h}\int\limits_{-h}^{h}dy\frac{\omega'v}{U}.$$ For fixed $v_*,$ $\nu,$ this average rapidly tends to zero as $h$ increases, because $U$ increases with $y,$ $v$ remains of order $v_*,$ while $\omega'$ decreases, for the Taylor length $\lambda$ appearing in the vorticity estimate $\omega\sim v_*/\lambda,$ and the length scale of the vorticity field (the Kolmogorov length $\lambda_0$) both grow with $h$ ($\lambda \sim h^{1/2},$ $\lambda_0 \sim h^{1/4}),$ so that $\omega'\sim h^{-3/4}$ in the bulk. Therefore, for sufficiently large $h,$ the terms involving $C$ in Eq.~(\ref{main2}) can be omitted. This equation thus takes the form
\begin{eqnarray}\label{main3}
\hat{H}\frac{d}{dy}\frac{1}{U}\left\langle \tilde{u}\hat{H}\frac{\partial^2 v}{\partial y\partial x} - \frac{\partial^2 \tilde{u}}{\partial x^2}v\right\rangle = - \frac{1}{U^2}\frac{d}{d y}\,U\!\left\langle v\hat{H}\frac{\partial^2 \tilde{u}}{\partial x^2} + \frac{\partial^2 \tilde{u}}{\partial y\partial x}v\right\rangle,
\end{eqnarray}\noindent where the continuity equation was also used to express all averages via the cross-correlation function (\ref{crosscor}).

As was already mentioned, this function depends on the streamwise coordinates only via their difference, and satisfies
\begin{eqnarray}\label{constv}
R(x,y;x,y) = - v^2_*.
\end{eqnarray}
\noindent In addition to that, continuity requires $\partial^2 R(x_1,y_1;x_2,y_2)/\partial x_1\partial y_2$ to be symmetric under the interchange $(x_1,y_1)\leftrightarrow (x_2,y_2),$
\begin{eqnarray}\label{symmetry}
\frac{\partial^2 R(x_1,y_1;x_2,y_2)}{\partial x_1\partial y_2} = - \left\langle  \frac{\partial \tilde{u}(x_1,y_1)}{\partial x_1}\frac{\partial \tilde{u}(x_2,y_2)}{\partial x_2}\right\rangle = \frac{\partial^2 R(x_2,y_2;x_1,y_1)}{\partial x_2\partial y_1}.
\end{eqnarray}
\noindent Otherwise, there are no general restrictions on $R(x_1,y_1;x_2,y_2),$ in particular, it may depend on $(y_1 + y_2)$ as well as on $(y_1 - y_2).$ The aim of the subsequent consideration is not to identify all possible functions $R(x_1,y_1;x_2,y_2)$ and $U(y)$ that satisfy Eq.~(\ref{main3}), but only a subclass thereof, which is general enough to include all three basic velocity profiles discussed in the Introduction, on the one hand, and simple enough to admit complete analysis, on the other. We therefore specialize to the simplest realization of the condition (\ref{constv}), wherein $R(x_1,y_1;x_2,y_2)$ depends on $y_1,y_2$ in the physical channel also via their difference only: $$R(x_1,y_1;x_2,y_2) = R(x_1-x_2,y_1-y_2;0,0) \equiv R(x_1-x_2,y_1-y_2).$$ Then $R(0,0) = - v^2_*,$ and the spectral density $\varrho$ can be conveniently introduced as
\begin{eqnarray}
R(x_1-x_2,y_1-y_2) = - \iint\limits_{-\infty}^{\infty}\frac{dk dq}{(2\pi)^2}\varrho(k,q)e^{ik(x_1-x_2) + iq(y_1-y_2)}, \quad y_1,y_2 \in (0,h).
\end{eqnarray}
\noindent As we deal with $h\gg \nu/v_*,$ decomposition into Fourier-integral with respect to the coordinate $y$ is used here instead of a Fourier series. According to the rule (\ref{crosscor1}), extension of $R$ over $y_1,y_2 \in (-h,h)$ reads
\begin{eqnarray}
R(x_1-x_2,y_1-y_2) = - \iint\limits_{-\infty}^{\infty}\frac{dk dq}{(2\pi)^2}\varrho(k,q)e^{ik(x_1-x_2) + iq(|y_1|-|y_2|)} \left\{\theta(y_1)\theta(y_2) - \theta(-y_1)\theta(-y_2)\right\}.\nonumber\\ \label{fourier}
\end{eqnarray}
\noindent Writing the density as $$\varrho(k,q) = \varrho_+(k,q) + \varrho_-(k,q),$$ where $\varrho_+(k,q)$ is even with respect to $k,$ and $\varrho_-(k,q)$ is odd, the symmetry property (\ref{symmetry}) requires $\varrho_+(k,q)$ be even also with respect to $q,$ and $\varrho_-(k,q)$ be odd:
\begin{eqnarray}\label{parity}
\varrho_+(k,q) &=& \varrho_+(-k,q) = \varrho_+(k,-q), \nonumber \\
\varrho_-(k,q) &=& -\varrho_-(-k,q) = - \varrho_-(k,-q).
\end{eqnarray}
\noindent Equation $R(0,0) = - v^2_*$ plays the role of normalization condition for $\varrho_+,$
\begin{eqnarray}\label{normalization}
\iint\limits_{-\infty}^{\infty}\frac{dk dq}{(2\pi)^2}\varrho_+(k,q) = v^2_*,
\end{eqnarray}
\noindent whereas Eq.~(\ref{symmetry}) implies a positivity requirement for $\varrho_-$:
\begin{eqnarray}\label{positivity}
\iint\limits_{-\infty}^{\infty}\frac{dk dq}{(2\pi)^2}\,kq\varrho_-(k,q) > 0.
\end{eqnarray}
\noindent In terms of the decomposition (\ref{fourier}), the averages appearing in Eq.~(\ref{main3}) read
\begin{eqnarray}\label{average1}
\left\langle \tilde{u}\hat{H}\frac{\partial^2 v}{\partial y\partial x}\right\rangle &=& \iint\limits_{-\infty}^{\infty}\frac{dk dq}{(2\pi)^2}\,\frac{kq\varrho_-(k,q)}{2h}\fint\limits_{0}^{h}d\tilde{y}e^{iq(|y|-\tilde{y})}
\cot\left\{\frac{\pi}{2h}(\tilde{y}-|y|)\right\}\chi(y) \equiv uHv_{xy}, \\
\left\langle\frac{\partial^2 \tilde{u}}{\partial x^2}v\right\rangle &=&
\iint\limits_{-\infty}^{\infty}\frac{dk dq}{(2\pi)^2}\,k^2\varrho_+(k,q) \chi(y)  \equiv u_{xx}v,\\
\left\langle v\hat{H}\frac{\partial^2 \tilde{u}}{\partial x^2}\right\rangle &=& \iint\limits_{-\infty}^{\infty}\frac{dk dq}{(2\pi)^2}\,\frac{k^2\varrho_+(k,q)}{2h}\fint\limits_{0}^{h}d\tilde{y}e^{iq(|y| - \tilde{y})}\cot\left\{\frac{\pi}{2h}(\tilde{y}-|y|)\right\} \equiv vHu_{xx}, \label{average3}\\
\left\langle\frac{\partial^2 \tilde{u}}{\partial y\partial x}v\right\rangle &=&
\iint\limits_{-\infty}^{\infty}\frac{dk dq}{(2\pi)^2}\,kq\varrho_-(k,q) \equiv u_{xy}v.\label{average4}
\end{eqnarray}
\noindent We observe that each of these functions depends on either $\varrho_+$ or $\varrho_-,$ that $u_{xy}v$ is a positive constant (Cf. (\ref{positivity})), whereas $u_{xx}v$ is constant in the physical and auxiliary channels separately. Rewritten in the abridged notation, Eq.~(\ref{main3}) takes the form
\begin{eqnarray}\label{main4}
\hat{H}\frac{d}{dy}\frac{[uHv_{xy} - u_{xx}v]}{U} = - \frac{1}{U^2}\frac{d}{d y}U[vHu_{xx} + u_{xy}v]\,.
\end{eqnarray}\noindent

\subsection{The near-wall asymptotic of the mean flow velocity}\label{nearwallasymp}

Given the functions defined by Eqs.~(\ref{average1})--(\ref{average4}), Eq.~(\ref{main4}) determines the mean velocity distribution $U(y)$ up to a multiplicative constant. The structure of $U(y)$ is thus essentially dependent on the details of correlation between $\tilde{u}$ and $v$ throughout the channel. It turns out, however, that the form of $U(y)$ for $y\ll h,$ that is its near-wall asymptotic exhibits remarkable universality: this asymptotic will be shown to depend only on a few integral characteristics of the cross-correlation spectrum. To begin with, we note that the small-$y$ asymptotics of the functions (\ref{average1}), (\ref{average3}) are determined by that of the integral
\begin{eqnarray}\label{integral}
I \equiv
\frac{1}{2h}\fint\limits_{0}^{h}d\tilde{y}e^{iq(|y|-\tilde{y})}
\cot\left\{\frac{\pi}{2h}(\tilde{y}-|y|)\right\}.
\end{eqnarray}\noindent The leading term of its small-$y$ asymptotic reads
\begin{eqnarray}\label{integral1}
I = \frac{1}{\pi}\ln\frac{2}{|qy|}\,.
\end{eqnarray}\noindent It is seen that the leading contribution is independent of $h$ and is singular at $y=0.$ This singularity is due to the pole of the cotangent at $\tilde{y} = |y|$ in the integrand of Eq.~(\ref{integral}). Recalling also that $u_{xy}v,$ $u_{xx}v$ are constants in each of the domains $y\in(-h,0),$ $(0,h),$ we observe that the combinations $[uHv_{xy} - u_{xx}v]$ and $[vHu_{xx} + u_{xy}v]$ appearing in Eq.~(\ref{main4}) have similar small-$y$ asymptotics, {\it viz.}, the sums of a logarithm and a constant. This suggests to write
\begin{eqnarray}\label{combination1}
uHv_{xy} - u_{xx}v &\equiv& [f_1(y) + \alpha]\chi(y),\\
vHu_{xx} + u_{xy}v &\equiv& [f_2(y) + \beta],\label{combination2}
\end{eqnarray}\noindent where $\alpha,\beta$ are as yet arbitrary constants, and to consider the following system
\begin{eqnarray}\label{main51}
\hat{H}\left\{\frac{d}{dy}\frac{f_1\chi}{U}\right\} &=& - \frac{1}{U^2}\frac{d\, \beta U}{d y}\,,\\
\hat{H}\left\{\frac{1}{U^2}\frac{d\, Uf_2}{d y}\right\} &=& \frac{d}{dy}\frac{\alpha\chi}{U}\,, \quad y\ne 0.\label{main52}
\end{eqnarray}\noindent If $U(y)$ is a solution of this system with $f_1,f_2,$ $\alpha,$ $\beta$ satisfying Eqs.~(\ref{combination1}), (\ref{combination2}), then taking the Hilbert transform of Eq.~(\ref{main52}) with the help of $\hat{H}^2=-1,$ and subtracting the result from Eq.~(\ref{main51}) shows that it is also a solution of Eq.~(\ref{main4}). On the other hand, $U(y)$ always exists such that it satisfies Eqs.~(\ref{main51}), (\ref{main52}) for {\it some} $f_1,f_2$ having the same logarithmic asymptotics for $y\ll h$ as the functions $uHv_{xy},$ $vHu_{xx},$ respectively. In fact, Eqs.~(\ref{main51}), (\ref{main52}) can be viewed as defining $f_1,f_2$ for a given $U(y)$ (this issue will be dealt with in detail in Sec.~\ref{reconstruction}). Therefore, as long as the assumption (A) is an acceptable approximation, the above system is equivalent to Eq.~(\ref{main4}), for then the bulk velocity fluctuations do not affect the near-wall behavior, so that the actual structure of the function $R(x,y)$ for $y\sim h$ is immaterial as to the form of $U(y)$ in the region $|y|\ll h.$ On the other hand, if (A) does not hold, transition from Eq.~(\ref{main4}) to the system (\ref{main51}), (\ref{main52}) implies that the cross-correlation function is not to be considered as prescribed, but treated self-consistently along with the mean velocity distribution. Anyway, the decoupling of functions $f_1,f_2,$ effected by replacing Eq.~(\ref{main4}) with Eqs.~(\ref{main51}), (\ref{main52}), is quite natural from the physical standpoint, because the even and odd components of the spectral density, $\varrho_+,$ $\varrho_-$ defining the leading (logarithmic) terms of the functions $f_2,f_1,$ respectively, are independent of each other, so that no cancelation between the corresponding terms in Eq.~(\ref{main4}) should be expected in general. But irrespective of this reasoning, and independently of whether or not (A) is applicable, the restriction imposed on the class of possible functions $R(x,y)$ by going over to the system (\ref{combination1}), (\ref{combination2}) is actually inessential, because all three types of the mean velocity profile -- the power, log, and power-log laws -- are still solutions to this system, as will now be proved. To this end, we note that Eqs.~(\ref{main51}), (\ref{main52}) allow $dU/d|y|$ to be expressed in two different ways
\begin{eqnarray}
\beta\frac{d}{d |y|}\frac{1}{U} &=& \chi\hat{H}\left\{\frac{d}{dy}\frac{f_1\chi}{U}\right\}\,,\nonumber\\
\frac{d}{d|y|}\frac{1}{U} &=& \frac{1}{\alpha}\hat{H}\left\{\frac{1}{U^2}\frac{d\, Uf_2}{d y}\right\}\,, \quad \alpha,y\ne 0.\nonumber
\end{eqnarray}\noindent These expressions can be combined to give
\begin{eqnarray}\label{main512}
\hat{H}\chi\hat{H}\frac{d}{dy}\frac{f_1\chi}{U} = - \frac{\beta}{\alpha U^2}\frac{d\, Uf_2}{d y}\,.
\end{eqnarray}\noindent According to Appendix A, the action of $\hat{H}\chi\hat{H}$ on a $2h$-periodic function $a(y)$ with zero spatial mean reads
\begin{eqnarray}\label{hchih1}
\left(\hat{H}\chi\hat{H}a\right)(y) = -\chi(y)a(y) + \frac{1}{\pi h}\int\limits_{-h}^{h}d\tilde{y} a(\tilde{y})\cot\left\{\frac{\pi}{2h}(y - \tilde{y})\right\}\ln\left|\frac{\tan (\pi\tilde{y}/2h)}{\tan (\pi y/2h)}\right|.
\end{eqnarray}\noindent This complicated expression greatly simplifies for $|y|\ll h$ and even $a(y).$ We first observe that the integrand is regular for $\tilde{y} = y.$ Therefore, integration over $\tilde{y} \sim y$ gives rise to a contribution which is $O(y/h)$ relative to the first term on the right of Eq.~(\ref{hchih1}). On the other hand, integration over $|\tilde{y}| \sim h$ is also $O(y/h)$ because the integrand becomes an odd function of $\tilde{y}$ on neglecting $y$ in the argument of the cotangent. As the function $(f_1\chi/U)'$ is even in $y,$ Eq.~(\ref{main512}) gives for $|y|\ll h$
$$\frac{d}{dy}\frac{f_1}{U} = \frac{\beta}{\alpha U^2}\frac{d\, Uf_2}{d y}\,,$$
or
\begin{eqnarray}\label{solution}
\left(\ln U\right)' = \frac{\alpha f'_1 - \beta f'_2}{\alpha f_1 + \beta f_2}\,.
\end{eqnarray}\noindent Since $f_{1,2}(y)$ are singular at $y=0,$ so are the solutions $U(y)$ to this equation. Depending on the value of the ratio $\beta/\alpha,$ we obtain the following classification of the near-wall velocity profiles:

\subsubsection{The logarithmic law}

According to Eqs.~(\ref{average1}), (\ref{integral1}) and (\ref{combination1}), the proportionality coefficient of the leading term in the function $f_1$ is equal to
$$\frac{1}{\pi}\iint\frac{dk dq}{(2\pi)^2}\,kq\varrho_-(k,q),$$ which is nonzero in view of the positivity property (\ref{positivity}) of $\varrho_-.$  Equation~(\ref{solution}) then shows that the case $\beta = 0,$ $\alpha\ne 0$ corresponds to solutions having logarithmic profile near the wall,
\begin{eqnarray}\label{loglaw}
U = \frac{v_*}{\varkappa}\ln\frac{|y|}{y_0}\,.
\end{eqnarray}\noindent Here the integration constant is written in the conventional form $v_*/\varkappa,$ where $\varkappa$ is the K$\acute{\rm a}$rm$\acute{\rm a}$n constant.
This constant is naturally left undetermined as we deal here only with the near-wall asymptotic of $U(y).$ The constant $y_0$ can be related to $\alpha$ and the characteristics of the cross-correlation spectrum:
\begin{eqnarray}\label{alpha}
\alpha =  - \iint \frac{dkdq}{(2\pi)^2}\left\{\frac{1}{\pi}kq\varrho_-\ln(|q|y_0/2) + k^2\varrho_+\right\}.
\end{eqnarray}\noindent As $U$ is positive, $\alpha$ must be chosen so as to place the point with the ordinate $y_0$ within the viscous layer.

\subsubsection{Power laws}\label{powerlaws}

When the parameters $\alpha,\beta$ satisfy
\begin{eqnarray}\label{powerlawcond}
\alpha\iint dkdq \,kq\varrho_- = - \beta\iint dkdq \,k^2\varrho_+
\end{eqnarray}\noindent the proportionality coefficient of the logarithm in the denominator of Eq.~(\ref{solution}) vanishes, and
the law of the wall becomes a power law:
\begin{eqnarray}\label{powerlaw}
U = A|y|^n,
\end{eqnarray}\noindent where the constant of integration $A$ is again arbitrary, while the exponent
\begin{eqnarray}\label{powerlawexp}
n= \frac{2\alpha\iint dkdq \,kq\varrho_-}{\iint dkdq \left\{kq\varrho_-\left[\alpha\ln(|q|y_0) - \beta\pi\right] + k^2\varrho_+\left[\beta\ln(|q|y_0) + \alpha\pi\right]\right\} + 4\pi^3(\alpha^2+\beta^2)}\,,
\end{eqnarray}\noindent $y_0 = O(\nu/v_*)$ being a constant of which $n$ is independent by virtue of Eq.~(\ref{powerlawcond}). $n$ must be positive for $U$ to grow away from the wall. For practical applications it will suffice to consider the case $n<1.$ Of particular interest is the limit of small $\alpha,\beta,$ because the power law can then be expected to be close to the log-law ($\beta=0$). Omitting terms quadratic in $\alpha,\beta$ in Eq.~(\ref{powerlawexp}), estimating the
integrals
\begin{eqnarray}
\iint dkdq \,k^2\varrho_+\ln(|q|y_0) &\approx& \ln\left(\frac{y_0}{\lambda_+}\right)\iint dkdq \, k^2\varrho_+, \nonumber\\
\iint dkdq \,kq\varrho_-\ln(|q|y_0) &\approx& \ln\left(\frac{y_0}{\lambda_-}\right)\iint dkdq \, kq\varrho_-,
\end{eqnarray}\noindent where $\lambda_+,\lambda_-$ are the characteristic length scales of the even and odd parts of the cross-correlation function, and taking into account condition (\ref{powerlawcond}) one finds, with a logarithmic accuracy,
\begin{eqnarray}\label{expapprox}
n \approx \frac{2}{\ln(\lambda_+/\lambda_-)}\,.
\end{eqnarray}\noindent According to Eq.~(\ref{normalization}), the value of the shear stress is determined by the even part of the cross-correlation function. Therefore, $\lambda_+$ is expected to be of order $h,$ because eddies of this size are most efficient in transporting momentum. On the other hand, $\varrho_-$ contributes not to the stress, but to the velocity gradients, and so the relevant length scale is the Kolmogorov length $\lambda_0 = h/{\rm Re}^{3/4}$ (observations\cite{jimenez2013,saddoughi1994} show that the characteristic length of the function $\langle (\partial \tilde{u}/\partial x)^2\rangle$ is actually about $40\lambda_0,$ but account of the proportionality factor is beyond the logarithmic accuracy). Thus,
\begin{eqnarray}\label{expapprox1}
n \approx \frac{8}{3\ln {\rm Re}}\ll 1\,.
\end{eqnarray}\noindent We conclude that the form of the power-law asymptotic characterized by sufficiently small $\alpha,\beta$ turns out to be universal, insensitive to the structure of the velocity correlations. Incidentally, that the exponent $n$ slowly decreases with the increase of the Reynolds number was known since the pioneering works of Prandtl and von K$\acute{\rm a}$rm$\acute{\rm a}$n, whereas the scaling $n\sim 1/\ln{\rm Re}$ was suggested much later in Ref.~\cite{barenblatt1993a} based on the analysis\cite{barenblatt1993b} of the same experimental data\cite{nikuradze} as was used earlier to verify the logarithmic law.\cite{karman}

\subsubsection{Power-log laws}

All other cases fall into the class of power-log laws:
\begin{eqnarray}\label{powerlog}
U = \frac{v_*}{\varkappa_p}\ln^p\frac{|y|}{y_0}\,.
\end{eqnarray}\noindent As before, the integration constant $\varkappa_p$ is arbitrary, whereas the constants $p$ and $y_0$ are related to the parameters $\alpha,\beta$
\begin{eqnarray}\label{powerp}&&
p = \frac{\iint dkdq\left(\alpha kq\varrho_- - \beta k^2\varrho_+\right)}{\iint dkdq\left(\alpha kq\varrho_- + \beta k^2\varrho_+\right)}\,, \\&&
\alpha^2+\beta^2 + \iint \frac{dkdq}{(2\pi)^2}\left\{\alpha k^2\varrho_+ - \beta kq\varrho_- + \left[\alpha kq\varrho_- + \beta k^2\varrho_+\right]\frac{1}{\pi}\ln\frac{y_0 |q|}{2}\right\} = 0\,.\label{powerab}
\end{eqnarray}\noindent As in the case of the log-law, parameters $\alpha,\beta$ must be such that $y_0$ is less than the viscous layer thickness. Also, $U$ grows away from the wall only for $p>0.$ Less trivial restrictions on the parameters will be established in Sec.~\ref{consequences}.

Finally, a note on the structure of the obtained solutions is in order. As we have seen, all three basic velocity distributions involve an arbitrary multiplicative constant -- the constant of integration of Eq.~(\ref{solution}). This is in fact a general property of solutions to the dispersion relation (\ref{main4}) which is homogeneous in $U.$ In principle, this arbitrary factor can be fixed by the normalization condition (\ref{norm}) once $U(y)$ satisfying Eq.~(\ref{main4}) is found for all $y\in(0,h).$ In this sense, Eqs.~(\ref{norm}), (\ref{main4}) allow complete solution to the problem of finding the mean velocity distribution for a given cross-correlation. However, they provide no relation between $V$ and the friction velocity $v_*$ that would allow one to fix the value of the K$\acute{\rm a}$rm$\acute{\rm a}$n constant $\varkappa$ (or $\varkappa_p$ in the power-log law (\ref{powerlog})).

\subsection{Reconstruction of the cross-correlation spectral density}\label{reconstruction}

Equation~(\ref{main4}) was used in Sec.~\ref{nearwallasymp} to determine possible profiles of the mean flow velocity near the wall. It is important, however, that the problem can be reversed to address the question as to what extent the mean velocity distribution in the channel affects velocity correlations. Viewing the problem from this standpoint reveals certain duality between these two objects, indicating that the question ``which velocity profile is the right one'' is probably not the right question to ask. As will now be proved, given a mean velocity distribution $U,$ one can always find a cross-correlation function $R$ such that $U,R$ satisfy Eq.~(\ref{main4}). In fact, there are infinitely many such $R$'s. To this end, we rewrite the system (\ref{main51}), (\ref{main52}) as
\begin{eqnarray}\label{main61}
f_1 &=& - \beta \chi U\hat{H}\left(\frac{1}{U} - c\right)\,,\\
f_2 &=& - \frac{\alpha}{U}\int\limits_{0}^{y}d\tilde{y} U^2(\tilde{y})\frac{d }{d\tilde{y}}\left(\hat{H}\frac{\chi}{U}\right)(\tilde{y}) \,,\label{main62}
\end{eqnarray}\noindent where the constant $c$ is the mean value of $1/U$ across the channel. Once $U(y)$ is given, these equations define the functions $f_{1,2}(y)$ for arbitrary $\alpha,\beta.$ Moreover, one can arrange things so as to fix the small-$y$ asymptotics of $f_{1,2}$ as desired. It is not difficult to see that in all three cases considered in Sec.~\ref{nearwallasymp}, the small-$y$ asymptotics of $f_{1,2}$ given by Eqs.~(\ref{main61}), (\ref{main62}) are logarithmic, $f_{1,2}(y)\sim \ln(|y|/h).$ The proportionality coefficients in these asymptotics can always be assigned any desired value by adjusting the parameters $\alpha,\beta,$ because according to Eqs.~(\ref{main61}),  (\ref{main62}), $f_1\sim \beta,$ $f_2 \sim \alpha.$ This proves the result referred to in the beginning of Sec.~\ref{nearwallasymp}.

Whether or not the asymptotics of $f_{1,2}$ are fixed, the cross-correlation spectral density can now be reconstructed as follows. One first differentiates and then Fourier-transforms Eqs.~(\ref{average1})--(\ref{average4}), (\ref{combination1}), (\ref{combination2}) with respect to $y,$ which yields two integral relations for the functions $\varrho_{\pm}(k,q)$
\begin{eqnarray}\label{inteq1}
\int\frac{dk}{2\pi}k^2\varrho_+(k,q) &=& - h\int\limits_{-h}^{h}d\tilde{y}f'_2(\tilde{y})
\sin\left(\frac{\pi\tilde{y}}{h}\right)\cos(q\tilde{y}), \\
\int\frac{dk}{2\pi}k\varrho_-(k,q) &=& - \frac{h}{q}\int\limits_{-h}^{h}d\tilde{y}f'_1(\tilde{y})
\sin\left(\frac{\pi\tilde{y}}{h}\right)\cos(q\tilde{y}). \label{inteq2}
\end{eqnarray}\noindent Since the dispersion relation holds only outside the viscous layer, $|y|\gg \nu/v_*,$ Eqs.~(\ref{inteq1}), (\ref{inteq2}) are valid only for $|q|\ll v_*/\nu.$ Thus, the form of $\varrho_{\pm}(k,q)$ at $q \gtrsim v_*/\nu$ remains arbitrary. On the other hand, the left hand sides of Eqs.~(\ref{combination1}), (\ref{combination2}), and the values of the constants $u_{xx}v,$ $u_{xy}v$ in particular, depend essentially on this form. This functional freedom in $\varrho_{\pm}$ can be used to fix, in infinitely many different ways, the constants that fell off from Eqs.~(\ref{combination1}), (\ref{combination2}) as a result of the $y$-differentiation. This proves the existence of $R$ satisfying the dispersion relation together with the given $U.$ It is clear that there is a continuum of such $R$'s, as the dependence on the argument $k$ of the spectral density also remains essentially arbitrary.

\subsection{Consequences of the assumption (A). Law of the wall in the limit ${\rm Re} \to \infty.$}\label{consequences}

The use of the assumption (A) has so far been mostly a matter of convenience: It has entailed no restriction on the near-wall flow structure, and its rather loose statement given in the Introduction called for no further specification. To explore nontrivial consequences of this assumption, we shall now make its statement more precise in the part concerning the mean velocity distribution:

(A$^\prime$) {\it For a given value of the shear stress, the leading term of the near-wall mean velocity asymptotic is independent of the cross-correlation between velocity fluctuations near the wall and in the bulk.}

In application to the system (\ref{main51}), (\ref{main52}) this means that the function $R(x,y)$ is required to be such that the functions $f_{1,2}(y)$ decrease away from the wall, so that the bulk contribution to the leading term of $U(y)$ can be neglected. This requirement on $R$ will be discussed in more detail later on, meanwhile the main consequence of (A$^\prime$) will be inferred, namely a reduction in the classification of the near-wall velocity profiles.

First of all, we note that the assumption just stated implies that the limit $h\to \infty$ can be taken in Eqs.~(\ref{main51}), (\ref{main52}), with all the functions involved replaced by the leading terms of their small-$y$ asymptotics, and the Hilbert operator written accordingly in the non-periodic form (\ref{hilbertinf}). Then it is readily seen that the power laws (\ref{powerlaw}) are inconsistent with (A$^\prime$). Consider, for instance, Eq.~(\ref{main52}). Substituting Eq.~(\ref{powerlaw}), the right hand side is seen to be proportional to $|y|^{-n-1}.$ However, setting $f_2(y) \sim  \ln|y|$ in the left hand side, and extracting the leading term yields an expression proportional to
$$\int\limits_{-\infty}^{\infty}\frac{d\tilde{y}\ln|\tilde{y}|}{|\tilde{y}|^{n+1}(\tilde{y} - y)} = 2\int\limits_{0}^{\infty}\frac{d\tilde{y}\ln \tilde{y}}{\tilde{y}^{n}(\tilde{y}^2 - y^2)}\,.$$ The formula\cite{prudnikov}
\begin{eqnarray}\label{integralf}
\int\limits_{0}^{\infty}\frac{d\tilde{y}\ln \tilde{y}}{\tilde{y}^{n}(\tilde{y}^2 - y^2)} = \frac{\pi}{4}|y|^{-n-1}\left\{\pi{\rm cosec}\frac{(1-n)\pi}{2} - 2\ln|y|\cot\frac{(1-n)\pi}{2}\right\}, \quad n<1,
\end{eqnarray}\noindent then shows that the leading term on the left is $\sim |y|^{-n-1}\ln|y|.$ We conclude that the mean velocity profiles characterized by the power-type near-wall asymptotics can be realized only in flows with sufficiently strong cross-correlation between velocity fluctuations near the wall and in the bulk. Evidently, such laws cannot be ``universal,'' characterized by parameters independent of the Reynolds number, because dependence on the flow properties at distances $|y|\sim h,$ hence on the value of $h$ itself is necessarily present. In particular, it is meaningless to speak about the limit $h \to \infty$ (that is, ${\rm Re} \to \infty$) in this situation.
On the contrary, a similar calculation shows that (A$^\prime$) can be accommodated by the log- and the power-log laws. The leading terms of the integrals appearing in Eqs.~(\ref{main51}), (\ref{main52}) can be extracted by writing
$$\int\limits_{-\infty}^{\infty}\frac{d\tilde{y}}{\ln^p(|\tilde{y}|/y_0)(\tilde{y} - y)} = \frac{2}{y}\int\limits_{0}^{\infty}\frac{d\eta}{\left[\ln\eta + \ln(|y|/y_0)\right]^p(\eta^2 - 1)} \approx \frac{-2p}{y\ln^{p+1}(|y|/y_0)}\int\limits_{0}^{\infty}\frac{d\eta \ln\eta}{(\eta^2 - 1)}\,,$$ and applying the formula (\ref{integralf}) with $n=0.$ The result is that the parameters $\alpha,\beta$ are no longer arbitrary, but related to the log exponent $p$ and the cross-correlation spectral density by
\begin{eqnarray}\label{alphabeta}
\alpha = - \frac{p+1}{2}\iint \frac{dkdq}{(2\pi)^2}k^2\varrho_+, \quad
\beta = \frac{p-1}{2}\iint \frac{dkdq}{(2\pi)^2}kq\varrho_-.
\end{eqnarray}\noindent These expressions are valid for $p=1$ as well as for $p>1.$ It is interesting to note that in the former case, combining the formula for $\alpha$ with Eq.~(\ref{alpha}) gives a simple equation for the length $y_0$ in the log-law (\ref{loglaw})
$$\iint \frac{dkdq}{(2\pi)^2}kq\varrho_-\ln(|q|y_0/2) = 0.$$ Within the logarithmic accuracy, this relation would imply $y_0 \sim h/{\rm Re}^{3/4}\,.$ However, this result should not be expected to be physically relevant. The point is that the logarithmic case is only a member of the continuous family of power-log solutions, and there is no reason {\it a priori} for $p$ to be equal unity {\it exactly}. On the contrary, possible constraints on the parameters are to be inferred from Eqs.~(\ref{alphabeta}), and (\ref{powerp}), (\ref{powerab}), and this turns out to be quite revealing as to the role of the case $\beta = 0.$ Equation (\ref{powerp}) becomes an identity on account of Eqs.~(\ref{alphabeta}), while Eq.~(\ref{powerab}) takes the form, with the logarithmic accuracy,
$$2\ln\frac{h}{y_0} - (p+1)\ln{\rm Re}^{3/4} + \frac{\pi}{2}(p^2 - 1)\frac{\iint dkdq\, k^2\varrho_+}{\iint dkdq\, kq\varrho_-} + \frac{\pi}{2}(p - 1)(p-3)\frac{\iint dkdq \, kq\varrho_-}{\iint dkdq \,k^2\varrho_+} = 0.$$ Recalling that the integral $\iint dkdq\,kq\varrho_-$ determines $\langle (\partial \tilde{u}/\partial x)^2\rangle\sim v^2_*/\lambda^2,$ whereas $\varrho_+$ is normalized on $v^2_*,$ $1/h$ being its characteristic wavenumber, the ratio of the integrals can be estimated as
$$\frac{\iint dkdq\, kq\varrho_-}{\iint dkdq\, k^2\varrho_+} \sim \frac{h^2}{\lambda^2}\,.$$ The Taylor length $\lambda$ scales with the Reynolds number as $\lambda \sim h/{\rm Re}^{1/2}.$ Therefore, the last term in the above equation is proportional to the large number ${\rm Re},$ and so cannot be compensated by the rest of the equation involving only logarithms of ${\rm Re},$ unless
$$p = 1 + f(\ln{\rm Re})/{\rm Re}\,,$$ where $f$ is close to the linear function for large values of its argument. The equation for $y_0$ thus reduces to
\begin{eqnarray}\label{approx}
\ln\frac{h}{y_0 {\rm Re}^{3/4}} = \frac{\pi f(\ln{\rm Re})}{2{\rm Re}}\frac{\iint dkdq \, kq\varrho_-}{\iint dkdq \,k^2\varrho_+}\,.
\end{eqnarray}\noindent We see that nothing actually can be said about the value of $y_0$ unless $f$ is known, and vice versa. What is important, however, is that this consideration brings to light peculiarity of the case $p=1$: the log exponent rapidly tends to unity as the Reynolds number increases, $(p-1)$ being roughly $O(\ln{\rm Re}/{\rm Re}).$ A more accurate estimate of this difference as well as its sign can be obtained by invoking the results of observations regarding the value of $y_0,$ which is usually written as $y_0 = a\nu/v_*.$ Though the value of $a$ noticeably varies from experiment to experiment, the variations seem to be around a ``universal'' value $a=0.13$ in a wide range of Reynolds numbers. The combination $v_*h/\nu \equiv {\rm Re}_*$ appearing in the argument of the logarithm on the left hand side of Eq.~(\ref{approx}) is related to ${\rm Re}=Vh/\nu$ roughly as ${\rm Re}_* = \varkappa\rm Re/\ln{\rm Re}.$ Therefore, this equation can be rewritten, with the same accuracy, as a formula for the log exponent:
\begin{eqnarray}\label{pminus1}
p = 1 + \frac{2}{\pi}\ln\left(\frac{\varkappa{\rm Re}^{1/4}}{a\ln{\rm Re}}\right)\frac{\iint dkdq \,k^2\varrho_+}{\iint dkdq \, kq\varrho_-}\,.
\end{eqnarray}\noindent The constants $\varkappa,a$ have been retained here because  $\ln({\rm Re}^{1/4}/\ln{\rm Re})$ is not large even for very large Reynolds numbers: it is only $1.25$ for ${\rm Re} = 10^7.$ According to (\ref{positivity}), the integral in the denominator is positive. As to the numerator, Eq.~(\ref{normalization}) generally requires positivity only for $\varrho_+$ integrated without the factor $k^2,$ but barring peculiar sign alternation in $\varrho_+,$ the numerator should normally also be positive. We then conclude that {\it $p$ tends to unity from above}.

It remains to specify conditions on the cross-correlation function under which (A$^\prime$) holds. As we know, the leading term of $U(y)$ can be identified as the limit of $U$ as $h\to \infty.$ It follows from Eqs.~(\ref{main51}), (\ref{main52}) that this term is independent of the bulk behavior of $f_{1,2}(y),$ that is (A$^\prime$) is met, provided that these functions decrease away from the wall. In view of the relations (\ref{average1}), (\ref{combination1}) and (\ref{average3}), (\ref{combination2}), this implies also certain restrictions on the $y$-dependence of the function $R(x,y),$ but these are strongly interrelated with the properties of the cross-correlations in the streamwise direction. This is clearly seen from Eqs.~(\ref{inteq1}), (\ref{inteq2}) showing that the functions $f_{1,2}(y)$ determine $q$-dependence of the integrals $\int dk k\varrho_-(k,q)$ and $\int dk k^2\varrho_+(k,q),$ rather than of the spectral density itself. Since no uniform conclusion regarding the $y$-dependence of $R(x,y)$ can be drawn under such circumstances, it is instructive to look at two opposite extreme cases, which are though both unrealistic, yet indicative as to the variety of possible cross-correlation functions consistent with (A$^\prime$). Of primary importance is the function $\varrho_+(k,q)$ as it determines cross-correlations in the wall-normal and streamwise directions, $R(0,y)$ and $R(x,0).$ In the case when its dependence on $k$ and $q$ factorizes, $\varrho_+(k,q) = \varrho^{(1)}_+(k)\varrho^{(2)}_-(q),$ it follows easily from Eq.~(\ref{inteq1}) that $R(x,y)\sim y f'_2(y),$ so that the cross-correlations decrease away from the wall roughly at the same rate as $f_2.$ However, in the opposite case of isotropic cross-correlation spectrum, $\varrho_+(k,q) = \varrho_+(k^2 + q^2),$ the factor $k^2$ in the integrand on the left of Eq.~(\ref{inteq1}) comes into play, bringing in an extra factor $y^{-2}$ when $|y|\gg |x|,$ so that $R$ scales with the distance to the wall as $R(x,y)\sim y^3f'_2(y).$
In principle, therefore, (A$^\prime$) can be met even in flows characterized by $R(x,y)$ growing with the distance as fast as $y^b,$ $b<2.$

\section{Discussion and conclusions}\label{conclusions}

Obtained under the quite restrictive condition of a two-dimensional flow, the above results appear nevertheless to admit lateral interpretation, in that the found qualitative features of the simplified configuration can be expected to be inherited by general three-dimensional flows. In the first instance, this concerns the existence of the power-log profiles (\ref{powerlog}) which were seen in Sec.~\ref{nearwallasymp} to appear naturally along with the logarithmic and power-law profiles as possible asymptotics of the near-wall mean velocity distribution. Found as solutions of the dispersion relation considered as an equation for the mean flow velocity, the power-log laws span a continuous family of which the logarithmic profile is just a member. Evidently, complexity of real flows can only enlarge the parameter space of this family, so that this state of things is to be expected to take place in general. Furthermore, the above considerations reveal the unique role played by the logarithmic case. The family continuity calls for special reasons to select any particular of its members, and the analysis of Sec.~\ref{consequences} gives a {\it dynamical} reason for the value $p=1$: the power-log exponent turns out to be a functional of the cross-correlation spectrum, which tends to unity as ${\rm Re} \to \infty,$ the difference $(p-1)$ being positive but small in practice.

This last result was obtained under the assumption (A$^{\prime}$) which turned out to be critical in the comparison of the power- and power-log families regarding the influence of the bulk flow on the near-wall velocity distribution. As was demonstrated in Sec.~\ref{consequences}, only the power-log profiles (including the case $p=1$) are consistent with the assumption that this influence is negligible, in the sense that the bulk velocity fluctuations do not affect the leading term of the near-wall mean velocity asymptotic. Conversely, the bulk effect on the flow near the wall must be sufficiently strong to sustain the mean velocity distribution obeying a power-law. The required strength of correlations is quantified in Sec.~\ref{reconstruction} in the form of Eqs.~(\ref{inteq1}), (\ref{inteq2}) which represent integral equations for reconstructing the cross-correlation spectrum that gives rise to the desired velocity profile. This result shows that from the purely theoretical standpoint, the question ``which velocity profile is the right one'' is a bit of a red herring, because given a mean velocity distribution, one can always find a cross-correlation function such as to satisfy the dispersion relation. There is little doubt that this conclusion also extends to the general three-dimensional case, but the question of whether the flow obeying given law is practically realizable can presently be answered only by the experiment. As mentioned in the Introduction, this issue is still a matter of debate, and it may well happen that flows with moderately large Reynolds numbers do obey the power law as the experiment seems to indicate, switching eventually to the log- or power-log law as the Reynolds number increases.

The found form of the power-law velocity profile deserves further discussion. The small-value limit of the exponent $n$ in Eq.~(\ref{powerlaw}) turned out to be a universal function of the Reynolds number: $n = 8/(3\ln{\rm Re}).$ Both the inverse proportionality to $\ln{\rm Re}$ and universality of its coefficient were inferred in Ref.~\cite{barenblatt2014} as the consequences of the so-called principle of vanishing viscosity, stating that the mean velocity gradient tends to a finite limit as $\nu \to \infty.$ This principle was suggested to replace the much stronger requirement that the mean velocity gradient be independent of $\nu,$ which leaves the logarithmic law as the only possibility for the near-wall velocity asymptotic. On the contrary, based on the analysis of extensive experimental data, the authors of Ref.~\cite{barenblatt2014}  insist on the practical relevance of the power law, and we see that the results of Sec.~\ref{powerlaws} lend purely theoretical support to the principle of vanishing viscosity.

The proportionality constant in the formula $n \sim 1/\ln{\rm Re}$ was inferred\cite{barenblatt2014} from the experiment to be $3/2$ instead of the calculated $8/3,$ and this leads us to the question of quantitative applicability of the obtained results. The two major simplifications employed in the analysis are the disregard of the memory effects and of the flow three-dimensionality. While the former appears to be a natural approximation in a fast flow (Cf. Sec.~\ref{asymptoticexp}), the latter most likely brings quantitative changes. These changes are difficult to assess without a three-dimensional generalization of the dispersion relation at hand, and the following comments are aimed merely on identifying the way the flow three-dimensionality is to be expected to show itself in the present context. Perhaps the most important feature lost upon restricting the flow to two dimensions is the mechanism of vortex stretching, and so it is a question of principle whether its inclusion into the framework of the dispersion relation is crucial. It should be clear from the considerations of Sec.~\ref{derivation} that neither the properties of the turbulent energy cascade, nor the way it actually develops (in particular, whether it is normal or inverse) is important for the present approach. Therefore, as the process driving the energy cascade, the vortex stretching should not be important either, and this conclusion is further substantiated by the following simple argument. To allow gradients in the flow velocity along the vortex-lines, consider the simplest generalization of the planar flow -- a three-dimensional flow described by the velocity field $\bm{v} = \bm{v}(x,y,t)$ in which all three velocity components depend only on two coordinates $x,y.$ All three components of vorticity are now generally nonzero, and by choosing an appropriate function $v_z(x,y)$ one can prescribe a fairly arbitrary distribution of the vortex stretch at any given instant. Subsequently, $v_z$ is transported as a passive scalar, but introducing a body force $f(x,y,t)$ acting in the $z$-direction, $v_z$ can be kept under control or even made stochastic, and such will be also the vortex stretching. In this situation, neither the velocity circulation, nor the $x,y$-components of vorticity are conserved, but $\omega_z$ is, and it is not difficult to see that all of the above results remain in force. Therefore, it is through the fluctuations in $\omega_z,$ rather than the vortex stretching, that the three-dimensionality of real flows is to be expected to affect these results in the first instance.

At last, a note on the minor simplification used in the analysis of the dispersion relation is in order, namely that of homogeneity of the cross-correlation function in the wall-normal direction. This is a purely technical assumption that allows one to write all results in an explicit analytical form, but it can be easily lifted if necessary. In particular, it is inconsequential regarding the classification of the near-wall mean velocity asymptotics given in Sec.~\ref{nearwallasymp}. In fact, in the case when the cross-correlation function depends not only on $(y_1-y_2)$ but also on $(y_1 + y_2),$ the quantities $\left\langle v\partial^2 \tilde{u}/\partial x^2\right\rangle$ and $\left\langle v \partial^2 \tilde{u}/\partial y\partial x \right\rangle$ in Eq.~(\ref{main3}) are no longer constant, yet bounded functions of $y,$ whereas those involving the Hilbert operator are still logarithmically divergent. Therefore, the analysis of Sec.~\ref{analysis} remains in force with the understanding that $y$ is small compared to the characteristic length associated with the dependence on $(y_1+y_2).$ Incidentally, the assumption that the flow is generated by the motion of the wall $y=h$ becomes useless on inclusion of the dependence of $R(x_1,y_1;x_2,y_2)$ on $(y_1+y_2).$ Yet, the derivation given in Sec.~\ref{derivation} proceeds in the same way and yields the dispersion relation in the same form whatever the flow structure in the bulk, provided that the turbulent flow is fully developed and homogeneous in the streamwise direction. Its local consequences such as the classification of the near-wall velocity profiles remain unchanged, whereas the global results such as the formulas Eq.~(\ref{inteq1}), (\ref{inteq2}) for reconstructing the cross-correlation spectral density are not universal and would require modification specific to each particular flow. At present, however, it would be premature to speculate on these modifications, because velocity correlations has been studied in far less detail than the mean flow.

\acknowledgments{This study was partially supported by RFBR, research project No. 13-02-91054~a.}

\begin{appendix}

\section{}

To derive the formula (\ref{hchih1}), we first write $\hat{H}\chi\hat{H}$ longhand using the definition (\ref{hilbert})
\begin{eqnarray}
\left(\hat{H}\chi\hat{H}a\right)(y) = \frac{1}{(2h)^2}\fint\limits_{-h}^{h}dy_1\chi(y_1) \cot\left\{\frac{\pi}{2h}(y_1 - y)\right\}\fint\limits_{-h}^{h}dy_2a(y_2)\cot\left\{\frac{\pi}{2h}(y_2 - y_1)\right\}.\nonumber
\end{eqnarray}\noindent The integral over $y_1$ cannot be taken directly by changing the order of integration, because of the point $y_1=y_2=y$ where the poles of the cotangents merge. To avoid this singularity, it is sufficient to rewrite the right hand side, for $y>0,$ as
\begin{eqnarray}
\frac{1}{(2h)^2}\fint\limits_{-h}^{h}dy_1\cot\left\{\frac{\pi}{2h}(y_1 - y)\right\}\fint\limits_{-h}^{h}dy_2a(y_2)\cot\left\{\frac{\pi}{2h}(y_2 - y_1)\right\} \nonumber\\ - \frac{2}{(2h)^2}\fint\limits_{-h}^{0}dy_1\cot\left\{\frac{\pi}{2h}(y_1 - y)\right\}\fint\limits_{-h}^{h}dy_2a(y_2)\cot\left\{\frac{\pi}{2h}(y_2 - y_1)\right\}.\nonumber
\end{eqnarray}\noindent The first term here is nothing but $(\hat{H}\hat{H}a)(y) = - a(y).$ Changing now the order of integration in the second, integrating over $y_1$ with the help of the identity
$$\cot x\cot y = -1 + \cot(x - y)(\cot y -\cot x),$$ and recalling the condition $\int_{-h}^{h} dya(y) = 0,$ one finds
\begin{eqnarray}
\left(\hat{H}\chi\hat{H}a\right)(y) = - a(y) + \frac{1}{\pi h}\int\limits_{-h}^{h}d\tilde{y} a(\tilde{y})\cot\left\{\frac{\pi}{2h}(y - \tilde{y})\right\}\ln\left|\frac{\tan (\pi\tilde{y}/2h)}{\tan (\pi y/2h)}\right|.\nonumber
\end{eqnarray}\noindent Combined with the result of a similar calculation for $y<0,$ this is Eq.~(\ref{hchih1}).

\end{appendix}

~


\vspace{-1cm}
\begin{thebibliography}{}

\bibitem{jimenez2013}
J.~Jim$\acute{\rm e}$nez, ``Near-wall turbulence,'' Phys.~Fluids {\bf 25}, 101302 (2013).

\bibitem{schultz2013}
M.~P.~Schultz and K.~A.~Flack, ``Reynolds-number scaling of turbulent channel flow,'' Phys.~Fluids {\bf 25}, 025104 (2013).

\bibitem{frisch1995}
U.~Frisch, {\it Turbulence. The legacy of A.N.~Kolmogorov} (Cambridge University Press,1995).

\bibitem{karman}
Th.~von~K$\acute{\rm a}$rm$\acute{\rm a}$n, ``Mechanische \"{A}hnlichkeit und turbulenz,'' Nachrichten der Akademie der Wissenschaften G\"{o}ttingen, Math.-Phys. Klasse, 58-76 (1930). [Engl. Trans.: ``Mechanical similitude and turbulence,'' Technical memorandum N611, National Advisory Committee for Aeronautics (Washington, 1931)].

\bibitem{monty2005}
J.~P.~Monty, ``Developments in smooth wall turbulent duct flows,'' Ph.D. dissertation (University of Melbourne, 2005).

\bibitem{zanoun2009}
E.-S.~Zanoun, H.~Nagib, and F.~Durst, ``Refined $c_f$ relation for turbulent channels and consequences for high-{\it Re} experiments,'' Fluid Dyn. Res. {\bf 41}, 021405 (2009).

\bibitem{hultmark2012}
M.~Hultmark, M.~Vallikivi, S.~C.~C.~Bailey, and A.~J.~Smits, ``Turbulent pipe flow at extreme Reynolds numbers,'' Phys. Rev. Lett. {\bf 108}, 094501 (2012).

\bibitem{degraaff2000}
D.~B.~DeGraaff and J.~K.~Eaton, ``Reynolds-number scaling of the flat-plate turbulent boundary layer,'' J. Fluid Mech. {\bf 422}, 319$-$346 (2000).

\bibitem{delalamo2004}
J.~C.~del~$\acute{\rm A}$lamo, J.~Jim´enez, P.~Zandonade, and R.~D.~Moser, ``Scaling of the energy spectra of turbulent channels,'' J. Fluid Mech. {\bf 500}, 135$-$144 (2004).

\bibitem{jimenez2008}
J.~Jim$\acute{\rm e}$nez and S.~Hoyas, ``Turbulent fluctuations above the buffer layer of wall-bounded flows,'' J. Fluid Mech. {\bf 611}, 215$-$236 (2008).

\bibitem{monty2009}
J.~P.~Monty, N.~Hutchins, H.~C.~H.~Ng, I.~Marusic, and M.~S.~Chong, ``A comparison of turbulent pipe, channel and boundary layer flows,'' J. Fluid Mech. {\bf 632}, 431$-$442 (2009).

\bibitem{barenblatt1993a}
G.~I.~Barenblatt, ``Scaling laws for fully developed turbulent shear flows. Part 1. Basic hypotheses and analysis,'' J. Fluid Mech. {\bf 248}, 513$-$520 (1993).

\bibitem{oberlack2001}
M.~Oberlack, ``A unified approach for symmetries in plane parallel turbulent shear flows,'' J. Fluid Mech. {\bf 427}, 299$-$328 (2001).

\bibitem{monin}
A.~S.~Monin and A.~M.~Yaglom, {\it Statistical fluid mechanics} (MIT Press, 1971).

\bibitem{lumley}
H.~Tennekes and J.~L.~Lumley, {\it A first course in turbulence} (MIT Press, 1972).

\bibitem{landau}
L.~D.~Landau and E.~M.~Lifshitz, {\it Hydrodynamics}, 2nd edn. (Pergamon, 1987).

\bibitem{kazakov1}
K.~A.~Kazakov, ``Exact equation for curved stationary flames with arbitrary gas expansion,'' Phys.~Rev.~Lett. {\bf 94}, 094501 (2005).

\bibitem{kazakov2}
K.~A.~Kazakov, ``On-shell description of stationary flames,'' Phys.~Fluids {\bf 17}, 032107 (2005).

\bibitem{jerk1}
H.~El-Rabii, G.~Joulin, and K.~A.~Kazakov, ``Nonperturbative approach to the nonlinear dynamics of two-dimensional premixed flames,'' Phys.~Rev.~Lett. {\bf 100}, 174501 (2008).

\bibitem{jerk2}
G.~Joulin, H.~El-Rabii, and K.~A.~Kazakov, ``On-shell description of unsteady flames,'' J.~Fluid~Mech. {\bf 608}, 217$-$242 (2008).

\bibitem{jerk4}
H.~El-Rabii, G.~Joulin, and K.~A.~Kazakov, ``Premixed flame propagation in channels of varying width,'' SIAM~J.~Appl.~Math. {\bf 70}, 3287 (2010).

\bibitem{jerk3}
H.~El-Rabii, G.~Joulin, and K.~A.~Kazakov, ``Stability analysis of confined V-shaped flames in high-velocity streams,'' Phys.~Rev. E {\bf 81}, 066312 (2010).

\bibitem{saddoughi1994}
S.~G.~Saddoughi and S.~V.~Veeravali, ``Local isotropy in turbulent boundary layers at high Reynolds numbers,'' J. Fluid Mech. {\bf 268}, 333$-$372 (1994).

\bibitem{barenblatt1993b}
G.~I.~Barenblatt and V.~M.~Prostokishin, ``Scaling laws for fully developed turbulent shear flows. Part 2. Processing of experimental data,'' J. Fluid Mech. {\bf 248}, 521$-$529 (1993).

\bibitem{nikuradze}
J.~Nikuradze, ``Gesetzm\"{a}ssigkeiten der turbulenten Str\"{o}mung in glatten R\"{o}hren,'' VDI Forschungsheft, No. 356 (1932).

\bibitem{prudnikov}
A.~P.~Prudnikov, Yu.~A.~Brychkov, and O.~I.~Marichev, {\it Integrals and series}, Vol.1, Chapter 2.6 (Nauka, Moscow, 1981), in Russian.

\bibitem{barenblatt2014}
G.~I.~Barenblatt, A.~J.~Chorin, and V.~M.~Prostokishin, ``Turbulent flows at very large Reynolds numbers: new lessons learned,'' Physics$-$Uspekhi {\bf 57}, 250$-$256 (2014).


\end{thebibliography}
\end{document}